\documentclass[twocolumn]{aastex631}

\usepackage{hyperref}
\usepackage{amsmath,amstext}
\usepackage[figure,figure*]{hypcap}
\usepackage{amssymb}
\usepackage{nicefrac}
\usepackage{graphicx}

\usepackage{wrapfig}

\graphicspath{{./}{figures/}} 

\shorttitle{Massive Black Hole Merger Host Galaxy Morphologies}
\shortauthors{Bardati et al.}

\begin{document}

\title{Signatures of Massive Black Hole Merger Host Galaxies from Cosmological Simulations I: \\ Unique Galaxy Morphologies in Imaging}

\correspondingauthor{Jaeden Bardati}
\email{jbardati@caltech.edu}

\author[0009-0002-8417-4480]{Jaeden Bardati}
\affiliation{Department of Physics \& Astronomy, Bishop's University, Sherbrooke, QC J1M 1Z7, Canada}
\affiliation{Division of Physics, Mathematics and Astronomy, California Institute of Technology, Pasadena, CA 91125, USA}

\author[0000-0001-8665-5523]{John J. Ruan}
\affiliation{Department of Physics \& Astronomy, Bishop's University, Sherbrooke, QC J1M 1Z7, Canada}

\author[0000-0001-6803-2138]{Daryl Haggard}
\affiliation{McGill Space Institute and Department of Physics, McGill University, 3600 rue University, Montreal, Quebec, H3A 2T8, Canada}

\author[0000-0002-4353-0306]{Michael Tremmel}
\affiliation{School of Physics, University College Cork, Cork, Ireland}

\begin{abstract}
Low-frequency gravitational wave experiments such as the \emph{Laser Interferometer Space Antenna} and pulsar timing arrays are expected to detect individual massive black hole (MBH) binaries and mergers. However, secure methods of identifying the exact host galaxy of each MBH merger amongst the large number of galaxies in the gravitational wave localization region are currently lacking. We investigate the distinct morphological signatures of MBH merger host galaxies, using the Romulus25 cosmological simulation. We produce mock telescope images of 201 simulated galaxies in Romulus25 hosting recent MBH mergers, through stellar population synthesis and dust radiative transfer. Based on comparisons to mass- and redshift-matched control samples, we show that combining multiple morphological statistics via a linear discriminant analysis enables identification of the host galaxies of MBH mergers, with accuracies that increase with chirp mass and mass ratio. For mergers with high chirp masses ($\gtrsim$10$^{8.2}$ $M_\odot$) and high mass ratios ($\gtrsim$0.5), the accuracy of this approach reaches $\gtrsim$80\%, and does not decline for at least $\sim$1~Gyr after numerical merger. We argue that these trends arise because the most distinctive morphological characteristics of MBH merger and binary host galaxies are prominent classical bulges, rather than relatively short-lived morphological disturbances from their preceding galaxy mergers. Since these bulges are formed though major mergers of massive galaxies, they lead to (and become permanent signposts for) MBH binaries and mergers that have high chirp masses and mass ratios. Our results suggest that galaxy morphology can aid in identifying the host galaxies of future MBH binaries and mergers.
\end{abstract}

\keywords{black holes -- galaxies: structure -- gravitational waves -- N-body simulations -- radiative transfer}


\section{Introduction} \label{sec:intro}

Mergers of massive black holes (MBHs; with black hole masses of $M_\mathrm{BH} \gtrsim 10^5 M_\odot$) at the centers of galaxies are an inevitable outcome of hierarchical galaxy formation \citep{Begelman_1980, Volonteri_2003, Volonteri_2009}. As two galaxies merge, their central MBHs will sink towards the gravitational potential minimum due to dynamical friction, and form a bound binary \citep{Colpi_1999, Yu_2002, Pfister_2017}. Additional angular momentum loss from three body interactions with stars \citep[e.g.,][]{Quinlan_1996, Merritt_2004, Berczik_2006, Vasiliev_2015, Sesana_2015} and gas torques \citep[e.g.,][]{Armitage_2002, Escala_2004, Dotti_2007, Mayer_2007, Cuadra_2009, Chapon_2013, Fontecilla_2019} will cause the binary to harden until its angular momentum loss is eventually dominated instead by gravitational wave radiation. Eventually, the binary will coalesce, producing a gravitational wave chirp \citep{Wyithe_2003, Sesana_2004}. Although mergers of stellar-mass black holes have been directly detected in high-frequency gravitational waves by LIGO/Virgo \citep{Abbott_2016}, mergers of individual MBHs have yet to be detected.

Current and future low-frequency gravitational wave experiments will directly detect MBHs binaries, both before and during their coalescence. Pulsar timing array \citep[PTA;][]{Jenet_2004} experiments such as those in the International Pulsar Timing Array project \citep{Verbiest_2016} are forecast to reach sufficient sensitivities to detect gravitational waves from individual MBH binaries of mass $\gtrsim$10$^8 M_\mathrm{BH}$ in nearby galaxies \citep{Sesana2009, Ravi_2015, Mingarelli_2017, Kelley_2018}. The \emph{Laser Interferometer Space Antenna} \citep[\emph{LISA};][]{Amaro-Seoane_2017} will detect the gravitational wave chirp from MBH mergers of mass $ 10^4 M_\odot \lesssim M_\mathrm{BH} \lesssim 10^8 M_\odot$ at redshifts up to $z\lesssim10$ \citep{Sesana_2005, Banks_2022}. These detections will provide new insights on the formation and growth of MBHs over cosmic time \citep[for reviews, see][]{Burke-Spolaor_2019, Amaro-Seoane_2022}.

Identifying the exact host galaxy counterpart of MBH binaries and mergers detected in gravitational waves will be critical for a variety of science goals. For example, multi-messenger detections of MBH mergers can be used as standard sirens to constrain cosmological parameters, through combining luminosity distance information from the gravitational waves, and recessional velocities of the host galaxy counterparts \citep{Schutz_1986, Holz_2005}. Furthermore, telescope observations can unveil the environment of the MBH binaries and probe the structure of their surrounding accretion flows, before and after coalescence \citep[e.g.,][]{Armitage_2002, Cuadra_2009, Noble_2012, Duffell_2020}. Methods for identifying the host galaxies of MBH mergers will be needed to realize these science goals.

Once a MBH binary or merger has been detected in gravitational waves, identifying the host galaxy will be challenging. Since the gravitational waves provide luminosity distance and MBH mass information, selection cuts based on redshift and stellar mass (assuming empirical galaxy scaling relations) can be made on the large number of galaxies lying in the localization region. For MBH binaries detected by PTAs, sky localizations are expected to be of order $\sim$10$^{2-3}$~deg$^2$ \citep{Sesana_2010, Goldstein_2018}. \citet{Goldstein_2019} show that even with a sophisticated approach combining both redshift and mass cuts on galaxies lying within the 90$\%$ sky localization regions of MBH binaries detected by PTAs, the number of host galaxy candidates is still predicted to be $\sim$10$^2$ galaxies. In contrast, the sky localizations of MBH mergers detected by \emph{LISA} are expected to be significantly smaller, of order $\sim$10$^{1-2}$~arcmin$^2$ after the merger \citep{Mangiagli_2020}. However, \citet{Lops_2023} show that there will nevertheless be of order $\sim$10$^{2-3}$ candidate galaxies lying in the 90$\%$ error volume for mergers of relatively-massive MBHs ($\gtrsim$10$^{7}$~$M_\odot$) at low redshifts ($z \lesssim 2$), for which the host galaxies can be detectable with deep wide-field telescope imaging. This problem is especially severe for these mergers, because although they are detectable by \emph{LISA}, their gravitational wave emission is near the low-frequency edge of the \emph{LISA} sensitivity curve, which will result in low signal-to-noise ratio (SNR) detections and correspondingly large error volumes \citep[][]{Klein_2016, Katz_2018}. Finally, many \emph{LISA} multi-messenger science goals aiming to detect electromagnetic emission from MBH binaries require identifying the host galaxy many days {\it prior} to coalescence. At these epochs, the temporally-evolving error volume will be significantly larger \citep[$\sim$10$^2$~deg$^2$ at 1~day before merger for a 10$^{7}$~$M_\odot$ MBH binary at $z = 1$;][]{Mangiagli_2020}, and contain $\sim$10$^{5-6}$ candidate host galaxies \citep{Lops_2023}. Thus, sophisticated methods for identifying the host galaxies of MBH binaries will need to comb through a large number of candidate galaxies in their gravitational wave error volumes.

Many methods for identifying host galaxies have focused on detecting electromagnetic signatures from the gas-rich environment surrounding the MBH binary \citep[see recent review by][]{Bogdanovic_2022}, but these approaches have disadvantages and large uncertainties. For example, MBH binaries detectable by PTAs may have circumbinary accretion disks that produce light curve periodicities \citep[e.g.,][]{Dorazio_2015, Graham_2015, Kelley_2019, Charisi_2022}, while MBH mergers detectable by \emph{LISA} may be followed by a transient afterglow from the circumbinary disk due to mass loss and recoil of the merged MBH \citep[e.g.,][]{Milosavljevic_2005, Megevand_2009, ONeill_2009, Corrales_2010, Rossi_2010, Tanaka_2010}. However, these predicted signatures either have not been directly observed, or are not confidently associated with MBH binaries, and thus their efficacy in identifying the host galaxy counterpart is still unclear. Furthermore, MBH binaries may preferentially reside in dusty environments \citep{Koss_2018}, causing any electromagnetic signatures from the binary to be difficult to detect. Finally, the majority of MBH binaries are expected to reside in gas-poor environments, and thus may not have detectable active accretion \citep{Izquierdo-Villalba_2023a, Dong-Paez_2023}. For these reasons, alternative approaches that do not assume a gaseous environment surrounding the MBH binary warrant closer investigation.

In this study, we investigate an alternative approach to identifying MBH binary and merger host galaxies, based on galaxy morphological signatures in broadband imaging. These morphological signatures may be expected since the formation of MBH binaries is preceded by the merger of their host galaxies. However, we emphasize that the approach we take here does {\it not} explicitly assume that the host galaxies of MBH binaries and mergers display morphological evidence of galaxy mergers, but we instead compare morphological statistics of MBH binary and merger host galaxies to control samples in an agnostic way. To do this, we use the Romulus25 cosmological simulation, and identify a sample of simulated galaxies that host MBHs mergers. The medium-volume medium-resolution nature of Romulus25 simultaneously enables the implementation of a sub-grid model for MBH dynamics \citep{Tremmel_2015}, and produces a relatively large sample of host galaxies for more statistically significant conclusions. We perform stellar population synthesis and dust radiative transfer to create mock broadband ultraviolet (UV), optical, and infrared (IR) telescope images of these simulated host galaxies, and extract morphological statistics for comparison to a redshift- and mass-matched control sample of galaxies.

Previous studies of MBH binary host galaxies using cosmological simulations have suggested that distinct morphological signatures will no longer be detectable at the time of the physical MBH merger, because morphological disturbances from the preceding galaxy merger will have disappeared \citep[e.g.,][]{Volonteri_2020, DeGraf_2021, Izquierdo-Villalba_2023b}. \citet{DeGraf_2021} used the Illustris simulation to investigate the signatures of galaxy mergers in MBH merger host galaxies. They showed that the simulated MBH merger host galaxies do initially display disturbed morphologies, indicative of a recent galaxy merger. However, this morphological evidence for galaxy mergers disappears in a few hundred Myr after the numerical MBH merger, well before the physical MBH merger that eventually occurs below the resolution limit of the simulation, thus leaving no trace of the preceding galaxy merger when the MBH merger is detected in gravitational waves. \citet{Volonteri_2020} used the NewHorizon and Horizon-AGN simulations to also study the host galaxies of MBH mergers. \citet{Volonteri_2020} more carefully estimated the delay-time from dynamical friction and hardening of the binary before physical MBH merger. They also concluded that at the time of MBH physical merger when gravitational waves would be detected, the simulated host galaxies do not display obvious disturbed morphologies indicative of the preceding galaxy mergers. In contrast to these studies, we compare the morphologies of MBH merger host galaxies to control samples in an agnostic way, without explicitly searching for evidence of galaxy mergers. Our study is thus \emph{also} sensitive to morphological signatures that are more permanent, in addition to galaxy merger-induced disturbances that disappear. Furthermore, we train a linear discriminant analysis (LDA) predictor to \emph{combinations} of morphological statistics to distinguish MBH merger host galaxies from the control sample, which has previously been shown to hold more discriminating power in classifying galaxies based on morphology \citep[e.g.,][]{Nevin_2019}.

The outline of our paper is as follows. In Section~\ref{sec:sims}, we describe the simulation, and our selection of both a sample of galaxies hosting MBHs mergers and a control sample. In Section~\ref{sec:radtrans}, we describe our production of mock images of these galaxies using radiative transfer. In Section~\ref{sec:imanalysis}, we describe our measurements of morphological statistics and the linear discriminant analysis. In Section~\ref{sec:discussion}, we interpret our results in context of previous studies. We summarize and conclude in Section~\ref{sec:conclusions}.

\section{Cosmological Simulation} \label{sec:sims}

\subsection{Romulus25}\label{subsec:romulus}

We use Romulus25 \citep[][]{Tremmel_2017}, a 25~cMpc per side uniform, periodic volume cosmological simulation, for our analysis. The Romulus25 simulation is run with \texttt{ChaNGa} \citep[][]{Menon_2015}, a N-Body + Smooth Particle Hydrodynamics code that includes sub-grid physics models such as a cosmic UV background, low temperature metal line cooling, star formation, and blastwave supernova feedback. Romulus25 also introduces new implementations of MBH formation, growth, and dynamics, briefly described below. The simulation assumes a $\Lambda$CDM cosmology using $\Omega_0$ = 0.3086, $\Omega_\Lambda$ = 0.6914, $h$ = 0.67, and $\sigma_8$ = 0.8288, consistent with \citet{Planck_Collaboration_2016}. The simulation is run to $z=0$, where it broadly agrees with empirical scaling relations between halo mass, galaxy stellar mass, and MBH mass. 

In Romulus25, MBH formation is tied directly to the properties of the surrounding gas. To seed a MBH, a gas particle that was selected to form into a star particle is instead converted to a MBH if it satisfies the following empirically-derived criteria: low metallicity ($Z/Z_\odot < 3 \times 10^{-4}$), gas density 15 times that of the star formation threshold, and temperature between 9,500~K and 10,000~K. MBHs are always seeded with a seed mass of $M_{\text{BH}} = 10^6 M_\odot$, accreted from nearby gas particles. This method more realistically models MBH seeding at high redshift in comparison to common seeding methods that are tied directly to the halo mass and {\it a priori} assumptions of MBH halo occupation fractions.

After the MBHs are seeded, they grow via accretion of the surrounding gas at a rate estimated through a modified Bondi-Hoyle prescription. This prescription accounts for angular momentum support of the gas at resolved scales, avoiding any additional sub-grid physics assumptions or free parameters. The free parameters from this model of MBH accretion and feedback, along with the star formation parameters, are constrained through comparison of the galaxy properties with empirical scaling relations. MBHs are numerically merged once they become closer than two softening lengths from each other ($\lesssim$700~pc), and if their relative velocities are low enough to be considered gravitationally bound. Romulus25 merges the black holes by simple addition, giving a central MBH mass of $M_{\text{BH}} = M_1 + M_2$. For more simulation details, we refer to \citet[][]{Tremmel_2017}.

MBHs in Romulus25 are tracked to sub-kpc scales by using a sub-grid model of dynamical friction \citep{Tremmel_2015}, which has been shown to accurately describe the sinking of `wandering' MBHs \citep{Tremmel_2018}. This dynamical friction model is an improvement over the more common approach in large volume cosmological simulations in which the MBHs are simply repositioned to the galaxy potential minimum at each timestep, which causes the MBHs in mergers of galaxies to immediately coalesce without any sinking time. In contrast, our approach allows MBHs to decouple from their host galaxy dynamics, enabling us to track the MBH merger dynamics down to separations of $\lesssim$700~pc (2$\times$ the gravitational softening length) before numerically merging. Thus, Romulus25 is better suited to studying MBH binaries and mergers, because there is less of a time delay between MBH {\it numerical} mergers (those declared by the simulation when their separation is than two softening lengths) and {\it physical} mergers (when a gravitational wave chirp would be produced). We emphasize that although the resolution and black hole dynamics sub-grid model of Romulus25 decrease this delay-time in comparison to previous studies using larger-volume simulations, there will nevertheless remain a significant additional delay-time of $\gtrsim$1~Gyr driven primarily by three-body interactions and gas torques at $\sim$parsec scales (well below the simulation resolution limit), before gravitational wave emission becomes the dominant source of angular momentum loss. Due to the highly-uncertain nature of this delay-time, we do not explicitly estimate a precise delay-time for each merger. Instead, we will investigate how our results change as a function of delay-time (see Section \ref{subsec:timescales}).

To identify gravitationally-bound dark matter halos and sub-halos, as well as the baryonic matter associated with these structures in the simulation, we use the \texttt{Amiga Halo Finder} \citep{Knollmann_2009}. Romulus25 contains galaxies in the virial mass range $\sim$$10^{8-13}~M_\odot$, and has a mass resolutions of $3.39 \times 10^5~M_\odot$ for dark matter particles, and $2.12 \times 10^5~M_\odot$ for gas particles. These mass resolutions are higher than large-volume cosmological simulations such as Illustris \citep{Sijacki_2015} or Horizon-AGN \citep{Volonteri_2016}, and the Romulus25 force resolution is similar to the highest-resolution EAGLES runs \citep{Schaye_2014}. The dark matter particles are also oversampled, which is key for enabling the improved black hole dynamics \citep[][]{Tremmel_2015}. The majority of our analysis of Romulus25 are performed using the \texttt{Pynbody} \citep[][]{Pynbody} and \texttt{Tangos} \citep[][]{Tangos} Python packages.

\begin{figure}[!tb]
\centering
\includegraphics[width=0.47\textwidth, keepaspectratio, angle=0]{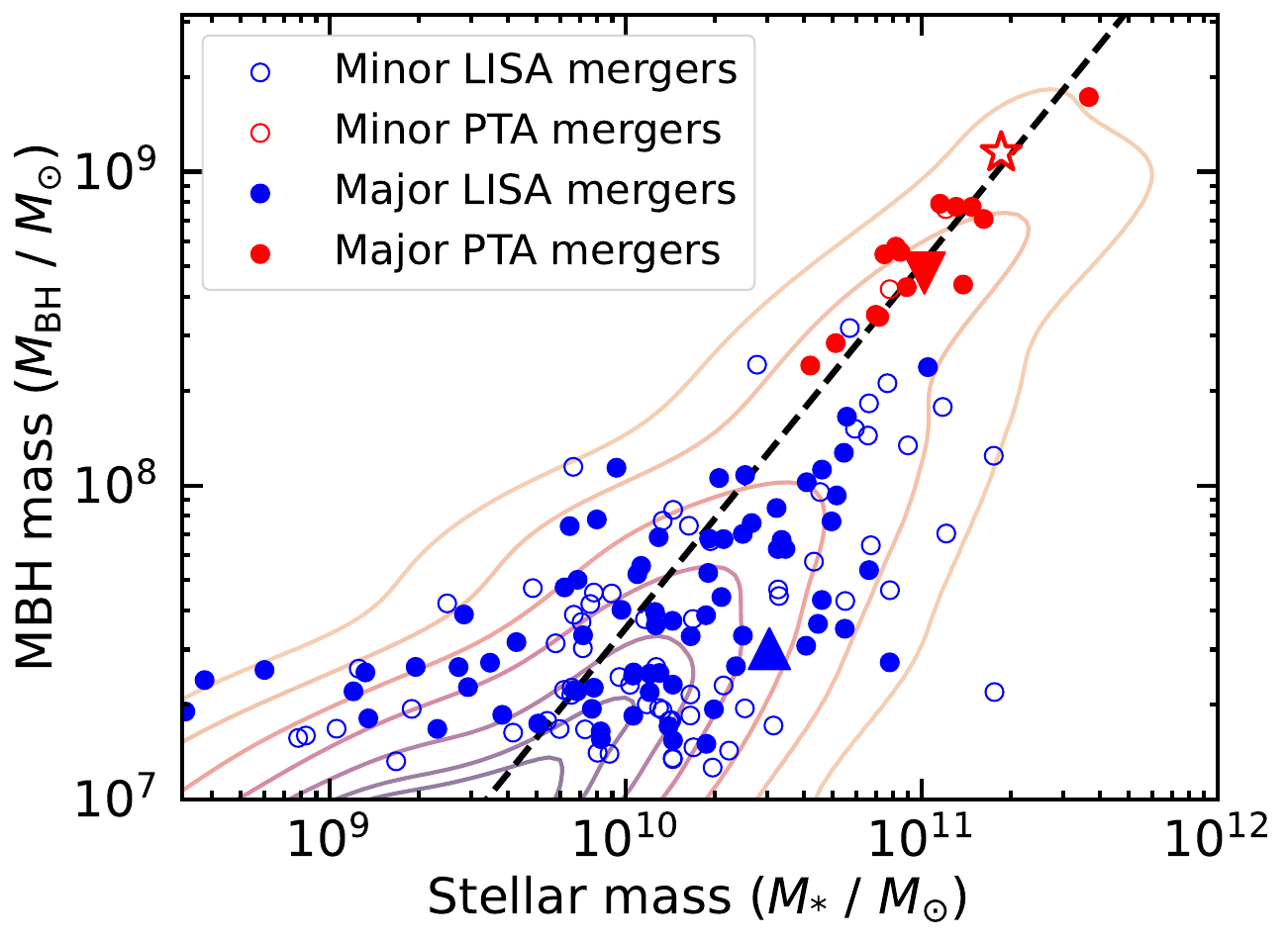}
\figcaption{The $M_\text{BH}-M_{*}$ relation for our MBH merger host sample, showing the major ($\nicefrac{M_2}{M_1} \geqslant 0.25$) vs.\ minor ($0.1 \leqslant \nicefrac{M_2}{M_1} < 0.25$) and \emph{LISA}-detectable ($M_{\text{chirp}} < 10^8 M_{\odot}$) vs. PTA-detectable ($M_{\text{chirp}} \geqslant 10^8 M_{\odot}$) MBH mergers. Each point represents the nearest host galaxy of a MBH merger event that has numerically merged within the last 400~Myr. The dashed line shows an empirical relation from \citet{Kormendy_Ho_2013}, for comparison. The contours show percentiles (55\% - 99.99\%) of all galaxies in Romulus25 and their corresponding central MBHs. The downward red triangle, upward blue triangle, and star red symbols indicate mergers that we use as examples in subsequent figures (a major PTA, major \emph{LISA}, and minor PTA merger, respectively). This shows that our sample of simulated MBH merger host galaxies conserves observed scaling relations that can be used for initial selection cuts on galaxies in the gravitational wave error volume.}
\label{fig:mbh-mstar-relation}
\end{figure}

\subsection{MBH Merger Host Galaxies Sample}\label{subsec:mergersampleselection}


We begin by selecting a sample of 201 MBH merger events in Romulus25 that host MBH numerical mergers. To do this, we select all MBH merger events in Romulus25 that occur at a redshift $z \leqslant 2$, with a mass ratio of $\nicefrac{M_2}{M_1} > 0.1$, where $M_1$ indicates the mass of the larger MBH (so $M_1 > M_2$), and with MBH masses $M_{\text{BH}} \gtrsim 10^7 M_\odot$. These cuts were chosen to select relatively massive galaxies at lower redshifts, for which wide-field telescope imaging can easily measure their morphologies. In Section~\ref{subsec:trends}, we show that our morphology-based approach is indeed most efficacious for mergers of the most massive MBHs with high mass ratios.

For each MBH merger in the sample, we identify its host galaxy at the nearest available simulation snapshot in time subsequent to the merger. We also track all galaxies in the sample for $\sim$1~Gyr after the numerical MBH merger. This enables us to also investigate the effects of the poorly-constrained delay-time between numerical merger and physical merger (see Section~\ref{subsec:timescales}). Since the time resolution of Romulus25 output snapshots at $z<2$ varies around 30 -- 300~Myr, we obtain approximately six epochs for each merger event, for a total of 1,057 simulated galaxies.

Figure~\ref{fig:mbh-mstar-relation} demonstrates that our sample of MBH merger host galaxies follows the $M_{\text{BH}}$ -- $M_{*}$ relation between central MBH mass and stellar mass at the time of numerical merger, consistent with other studies probing MBH mergers using cosmological simulations \citep[e.g.,][]{DeGraf_2021}, as well as the observed $M_{\text{BH}}$ -- $M_{*}$ relation of galaxies from \citet{Kormendy_Ho_2013}. In Figure~\ref{fig:mbh-mstar-relation}, we separate the sample into `major' and `minor' mergers, in which major mergers are defined to have mass ratios of $\nicefrac{M_2}{M_1} \geqslant 0.25$, while minor mergers have $0.1 \leqslant \nicefrac{M_2}{M_1} < 0.25$. Furthermore, we crudely label mergers as \emph{LISA}-detectable if they have chirp masses of $M_{\text{chirp}} \equiv {(M_1 M_2)^{3/5}} / {(M_1 + M_2)^{1/5}} < 10^8 M_{\odot}$, and PTA-detectable if $M_{\text{chirp}} \geqslant 10^8 M_{\odot}$. Since MBHs in Romulus25 are initially seeded with masses of $10^6 M_{\odot}$, we place no minimum limit on the MBH mass for \emph{LISA}-detectable mergers. For our MBH merger host galaxies sample, we also find that there is a relation between $M_{\text{chirp}}$ and $M_{\text{BH}}$, with the difference between the masses given by $\log(M_{\text{chirp}}) - \log(M_{\text{BH}}) = 0.5 \pm 0.1~(0.40 \pm 0.04)$ for major (minor) mergers. Thus, our sample of MBH merger hosts conserves observed scaling relations that can be used to place initial selection cuts on galaxies in the gravitational wave error volume. 

\subsection{Control Galaxy Sample}\label{subsec:controlsampleselection}

\begin{figure}[!tb]
\centering
\includegraphics[width=0.44\textwidth, keepaspectratio, angle=0]{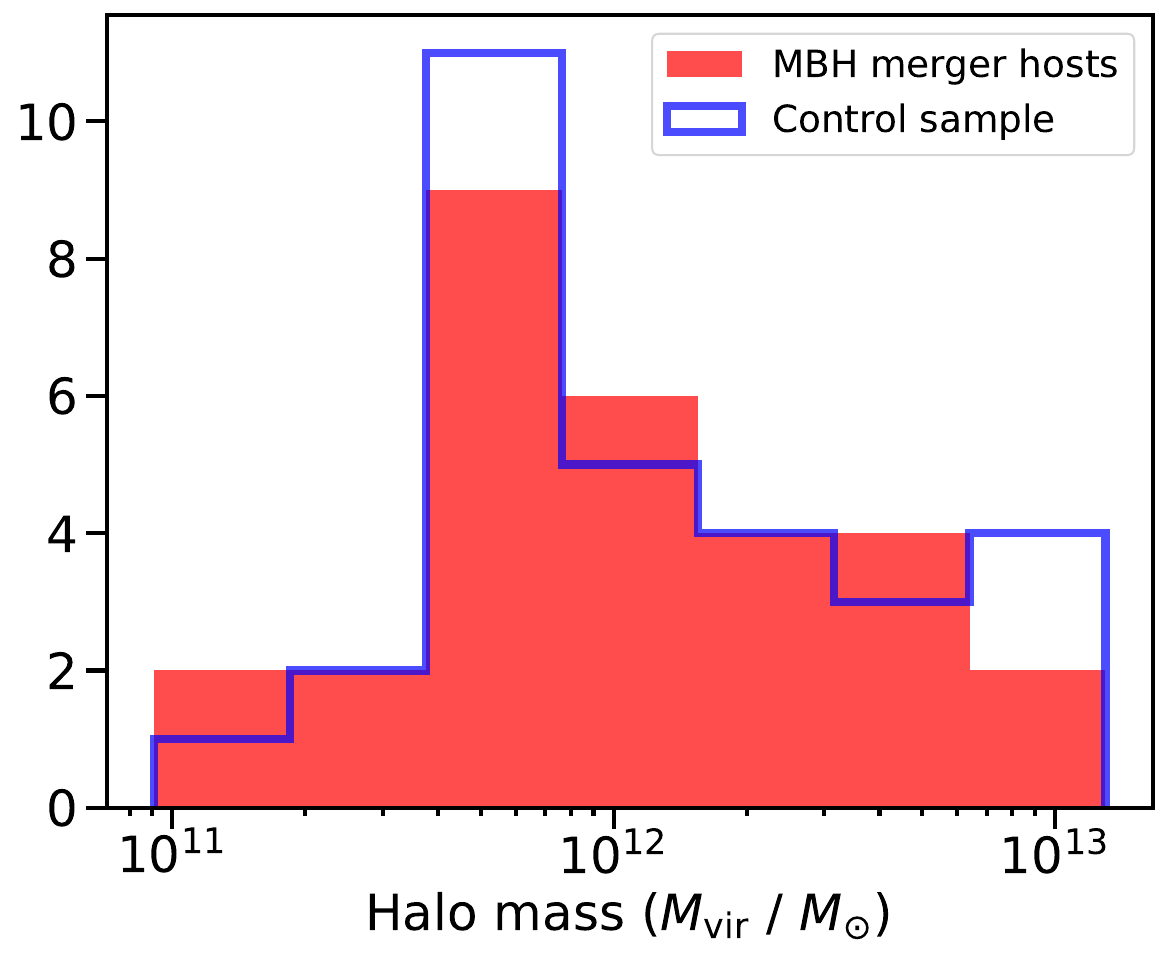}
\figcaption{An example mass histogram at a particular time bin ($0.31 < z < 0.34$), showing the mass distribution for this subsample of MBH merger host galaxies, as well as its corresponding redshift- and mass-matched control galaxies.}
\label{fig:example-time-bin-mass-dist}
\end{figure}

To quantify the accuracy of our approach in distinguishing between MBH merger host galaxies and non-merger hosts, we also create a control sample of simulated galaxies. To avoid biases stemming from galaxy mass and redshift in our results, we mass- and redshift-match our control sample, following a method similar to \citet[][]{DeGraf_2021}. We begin by binning by timestep with a minimum of 30 merger events per time bin, sufficient to produce a reasonable mass distribution in each time bin. The timestep bins typically span one or two timesteps ($\sim$100 Myr). We then mass bin the MBH merger host galaxies in each time bin. For each mass bin, we calculate a weight as the ratio of the number of MBH merger host samples to the number of non-merger galaxies in the bin. This weight is then applied to all non-merger galaxies in the bin. The resulting weighted distribution of non-merger galaxies is repeatedly sampled, obtaining a control sample of non-merger host galaxies that is mass-matched to the corresponding MBH merger host galaxies in each time bin. An example mass distribution in a time bin ($0.31 < z < 0.34$) is shown in Figure~\ref{fig:example-time-bin-mass-dist}, for both MBH merger hosts and the corresponding control sample. The collection of all time bins is our resulting mass- and redshift-matched control sample.

\section{Dust Radiative Transfer Simulations}\label{sec:radtrans}

\subsection{Powderday}\label{subsec:powderday}

To generate mock images, we use the \texttt{Powderday} \citep[][]{Narayanan_2021} dust radiative transfer software package. \texttt{Powderday} performs stellar population synthesis using Flexible Stellar Population Synthesis \citep[\texttt{FSPS};][]{Conroy_Gunn_2010}, 3D Monte Carlo dust radiative transfer using \texttt{Hyperion} \citep[][]{Robitaille_2011}, and uses \texttt{yt} \citep[][]{Turk_2011} to interface between various simulation data formats. The mock images of each galaxy in the merger-host and control sample are generated as follows. 

\begin{figure*}[t!]
\centering
\includegraphics[width=0.95\textwidth]{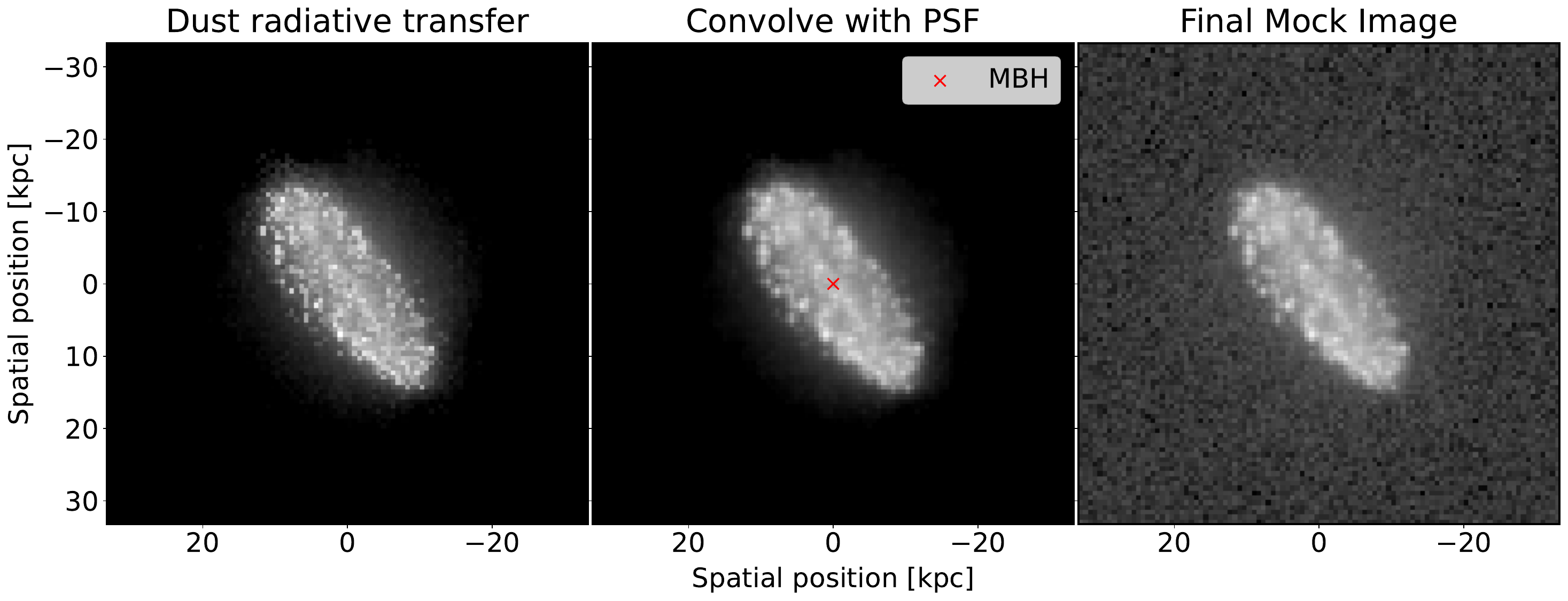}
\figcaption{The steps in processing the mock images to match telescope specifications. Starting with a \texttt{Powderday} dust radiative transfer image (left panel), we convolve with a $0\farcs1$ FWHM PSF (middle panel), and introduce a residual background noise (right panel). The final mock image is a \emph{CASTOR} UV image for a host galaxy that corresponds to MBH merger event marked with a red star from Figure~\ref{fig:mbh-mstar-relation}.}
\label{fig:matching-mock-images}
\end{figure*}

Using stellar population synthesis, stellar spectral energy distributions (SEDs) are generated for each star particle based on its age, metallicity and mass. We treat each star particle as a Simple Stellar Population, where the star formation history is a sharp burst with fixed metallicity and age. We use the \citet{Kroupa_2001} initial mass function, as well as the MIST evolutionary tracks \citep{Jieun_2016}. We use the MILES \citep{Sanchez_2006} spectral library \citep{Vazdekis_10} to obtain a spectrum from the single stellar population in the particle, in addition to Binary Population and Spectral Synthesis \citep[BPASS;][]{Eldridge_2017} to account for the evolution of binary systems. Since we are interested in the host galaxy rather than emission from MBH accretion, we do not include accretion disk emission in the SEDs. Finally, we do not include emission from nebular lines and Polycyclic Aromatic Hydrocarbon molecules in our SEDs, since these narrow emission lines have relatively weak contributions to our mock broadband images.

Using dust radiative transfer, the stellar SEDs are projected through the interstellar medium to generate a synthetic galaxy SED. \texttt{Hyperion} forms a dust-grid with the \citet{Draine_2003} Milky Way dust model with $R_V=3.1$ and $b_c=60 \text{ppm}$, and assumes a constant dust-to-metal ratio of 0.25 (consistent with \citealt{Li_2019}). Each source emits photon packets in a random direction and frequency based on their SED, which are propagated and attenuated throughout the grid. Ray tracing is then used to generate the resulting galaxy SED. 

We next convolve the resultant galaxy SED with UV, optical, and IR telescope filters to obtain mock surface brightness images in those bands. Specifically, since telescope follow-up in the localization regions of MBH mergers will likely be performed with wide-field space-based imaging, we use the \emph{Cosmological Advanced Survey Telescope for Optical and ultraviolet Research} \citep[\emph{CASTOR;}][]{Cote_2012} broadband UV and optical $g$-band filters, and the \emph{Nancy Grace Roman Telescope} \citep[\emph{Roman};][]{Spergel_2015} F146 broadband IR filter. The \emph{CASTOR} UV filter covers the approximate wavelength range $\sim$ 1740 -- 2830~\AA~ ($\lambda_{\text{eff}} \approx 2290$~\AA), the $g$-band covers $\sim$ 4000 -- 5520~\AA~($\lambda_{\text{eff}} \approx 4770$~\AA), while the \emph{Roman} F146 filter covers $\sim$ 0.92 -- 2.01~$\mu$m ($\lambda_{\text{eff}} \approx 1.473$~$\mu$m). Together, these filters cover a broad range of wavelengths to probe galaxy morphology.


\begin{figure*}[t]
\centering
\includegraphics[width=0.98\textwidth]{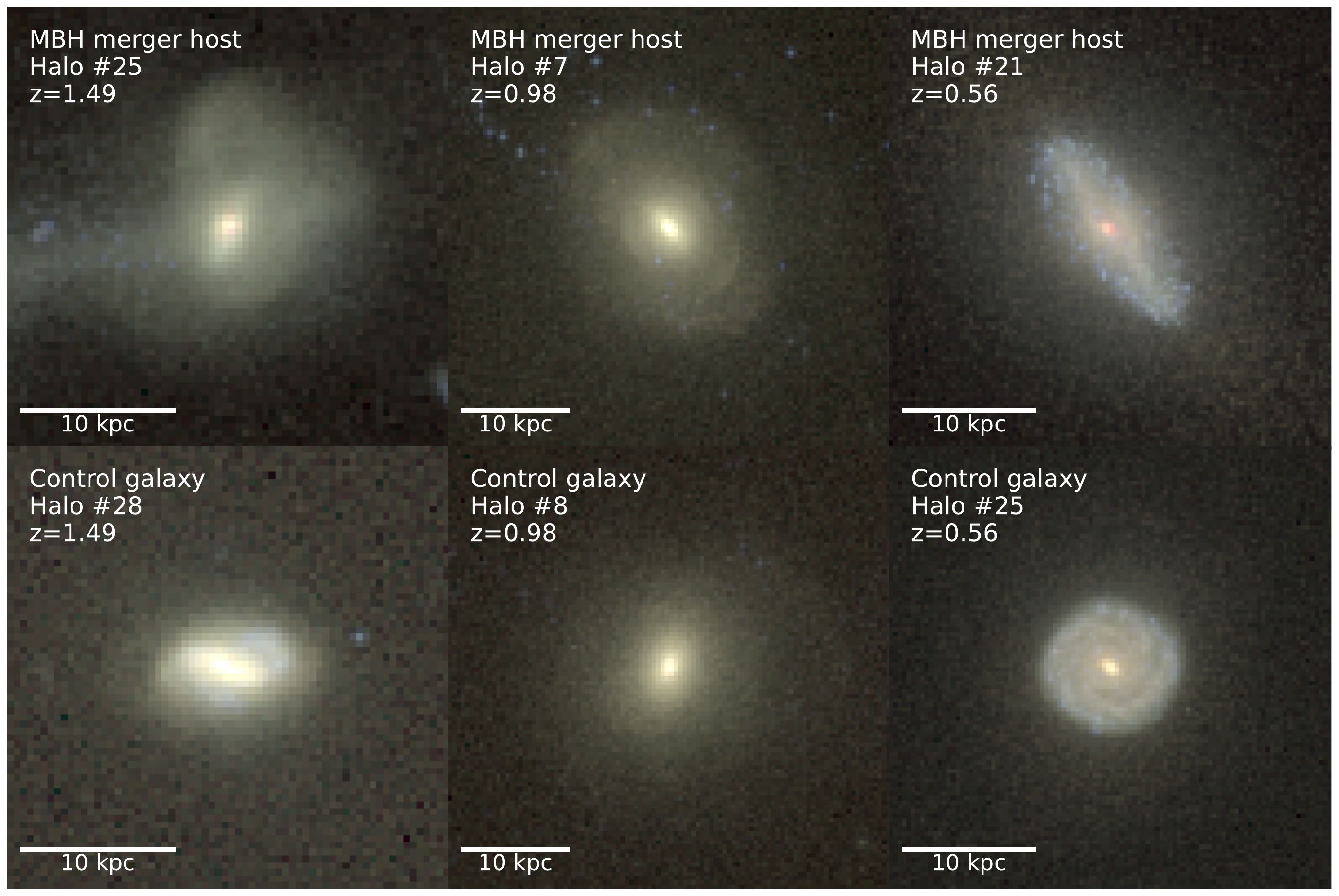}
\figcaption{A comparison of the 3-color images of simulated galaxies from the sample of MBH merger hosts (top images) and control sample (bottom images). From left to right, the MBH merger host images correspond to the upward triangle, downward triangle, and star MBH merger events shown in Figure~\ref{fig:mbh-mstar-relation}. The control galaxies chosen are matched in mass and redshift to the corresponding sample galaxies above them. We create these red-green-blue (RGB) visualizations using our three filters (\emph{Roman} IR, optical \textit{g}-band, and \emph{CASTOR} UV, respectively), through standard \citet{Lupton_2004} asinh scaling. We note that our analysis in Section~\ref{sec:imanalysis} are performed using images in each filter separately rather than these 3-color images.}
\label{fig:radiative-transfer-results}
\end{figure*}

\subsection{Matching Telescope Specifications}\label{subsec:telescopespecs}

To make our mock images more realistic, we choose spatial resolutions similar to those expected for \emph{CASTOR} and \emph{Roman}, convolve with a Point Spread Function (PSF), and add white noise, following a method similar to \citet{Nevin_2019} and \citet{sharma2021connection}. This process is illustrated for a example mock galaxy image in Figure~\ref{fig:matching-mock-images}, and we describe this process in more detail below. 

We first resize each image based on the redshift of the galaxy in Romulus25. We assume a pixel scale of $0\farcs1~\text{pixel}^{-1}$ (similar to that planned for \emph{CASTOR} and \emph{Roman}), and calculate the equivalent physical length (in kpc) for the given redshift. This corresponds to pixel length scales of approximately 400 -- 900~pc. Each image is clipped to 100 $\times$ 100 pixels, sufficient to contain the entire galaxy. The resultant  100 $\times$ 100 pixel images thus have varying physical lengths on each side, but a constant pixel scale of $0\farcs1~\text{pixel}^{-1}$.

We next convolve each image with a Gaussian PSF with a effective Full Width at Half Max (FWHM) of $0\farcs1$, using the \texttt{Astropy} package \citep{Astropy_2022}. This is also chosen to approximately match the PSF expected for \emph{CASTOR} and \emph{Roman}. 

Finally, we add a residual background noise to the image. To do this, we first find the mean flux ($f_\mathrm{eff}$) in a 1 pixel annulus with circular half-light radius ($R_\mathrm{eff}$) centered at the centroid of the image. Assuming a signal to noise $\mathrm{SNR}=100$ at $R_\mathrm{eff}$, we convert to approximate data counts by multiplying the flux image by the conversion factor $\mathrm{SNR}^2/f_\mathrm{eff}$. Although this conversion factor is idealized, we find that its exact value has little effect on the morphological measurements at large SNR. We then add white noise to each pixel by drawing from a Poisson distribution with $\lambda = \mathrm{SNR}$. We do not carefully consider instrumental gain, since the SNR in the image will be primarily determined by the exposure time of the observation in practice. Although our approach is simple, we find that the addition of this noise aids in regularization of our image segmentation, deblending, and morphological measurements in Section~\ref{subsec:mergerpredictors}.

To mitigate the effects of inclination on our measurements of galaxy morphological parameters, we produce mock images of each galaxy at a variety of different viewing angles. Specifically, we generate mock images of each galaxy in both the merger-host and control samples from four different isotropically-oriented viewing angles. For each galaxy, we first center the galaxy halo, and then generate mock images from the perspective of each of the four vertices of a regular tetrahedron centered around its halo. Since galaxies are assumed to be randomly oriented, the tetrahedron is also effectively randomly oriented, and thus the resulting mock images sample a variety of different viewing angles. This is supported by our Sérsic fitting analysis conducted on both mock image samples, revealing uniform distributions of galaxy orientations.

A random selection of the resultant mock images are displayed in Figure~\ref{fig:radiative-transfer-results}, from both our MBH merger host sample and the corresponding control sample.

\section{Analysis of the Mock Images}\label{sec:imanalysis}

\subsection{Morphology Measurements}\label{subsec:mergerpredictors}

Prior to extracting the morphological measurements, we first perform segmentation on the galaxies in the mock images to remove any satellite galaxies and define the boundaries of the primary galaxy. The segmentation assigns a label to every pixel in the image identifying the source to which it belongs (i.e. background noise, primary or satellite galaxy). A galaxy source is defined as being a minimum number of connected pixels whose flux values are above a threshold multiple of the background noise. Specifically, we create a segmentation map using a threshold of $10\sigma$ above the background with a minimum of 10 connected pixels, and regularized with a 5 pixel boxcar kernel. To account for galaxies in the process of merging or for satellites that lie along the line of sight of the particular viewing angle, we also perform deblending on this segmentation map. This is especially important for measurements that are sensitive to irregular galaxies, such as the Sérsic index. 

We extract seven different morphological measurements:  Gini coefficient, $M_{20}$, CAS parameters, shape asymmetry ($A_S$) and Sérsic index ($n$). We follow the definitions of these parameters from \citet{Lotz_2004} and \citet{Pawlik_2016}, and use the \texttt{StatMorph} package \citep[][]{Rodriguez-Gomez_2019}. We describe each of these measurements in turn below.

The Gini coefficient measures how evenly the galaxy flux is distributed, irrespective of the center of the galaxy. The Gini coefficient is high for bright single or dual nuclei and low for galaxies with more uniform surface brightnesses, ranging from 0 to 1. We use the equation for the Gini coefficient from \citet{Lotz_2004}:

    \begin{equation*}
        Gini = \frac{1}{|\Bar{X}|n(n-1)}\sum_i^n(2i-n-1)|X_i|,
    \end{equation*}

where $X_i$ is the flux value of the $n$ pixels in increasing order of brightness, and $\Bar{X}$ is the average flux value. 

The $M_{20}$ coefficient describes the concentration of light in a galaxy, also irrespective of the center of the galaxy. Specifically, the $M_{20}$ coefficient is defined as the second-order moment of the brightest 20\% of pixels:

\begin{equation*}
    M_{20} = \log_{10}\left(\frac{\sum_i^m M_i}{M_\text{tot}}\right) \text{for} \sum_i^m f_i < 0.2f_\text{tot},
\end{equation*}

where $f_i$ are the fluxes ordered from largest to smallest, $X_\text{tot}$ is the total flux of all pixels in the segmentation map, and $M_i = f_i\left((x_i-x_c)^2+(y_i-y_c)^2\right)$. The number of pixels summed over $m$ is defined such that the sum of the top $m$ brightest pixel fluxes are less than but closest in value to 20\% of the total flux. The center ($x_c$,$y_c$) is computed such that $M_\text{tot} \equiv \sum_i^n M_i$ is minimized.

\begin{figure*}[ht]
\centering
\includegraphics[width=0.99\textwidth]{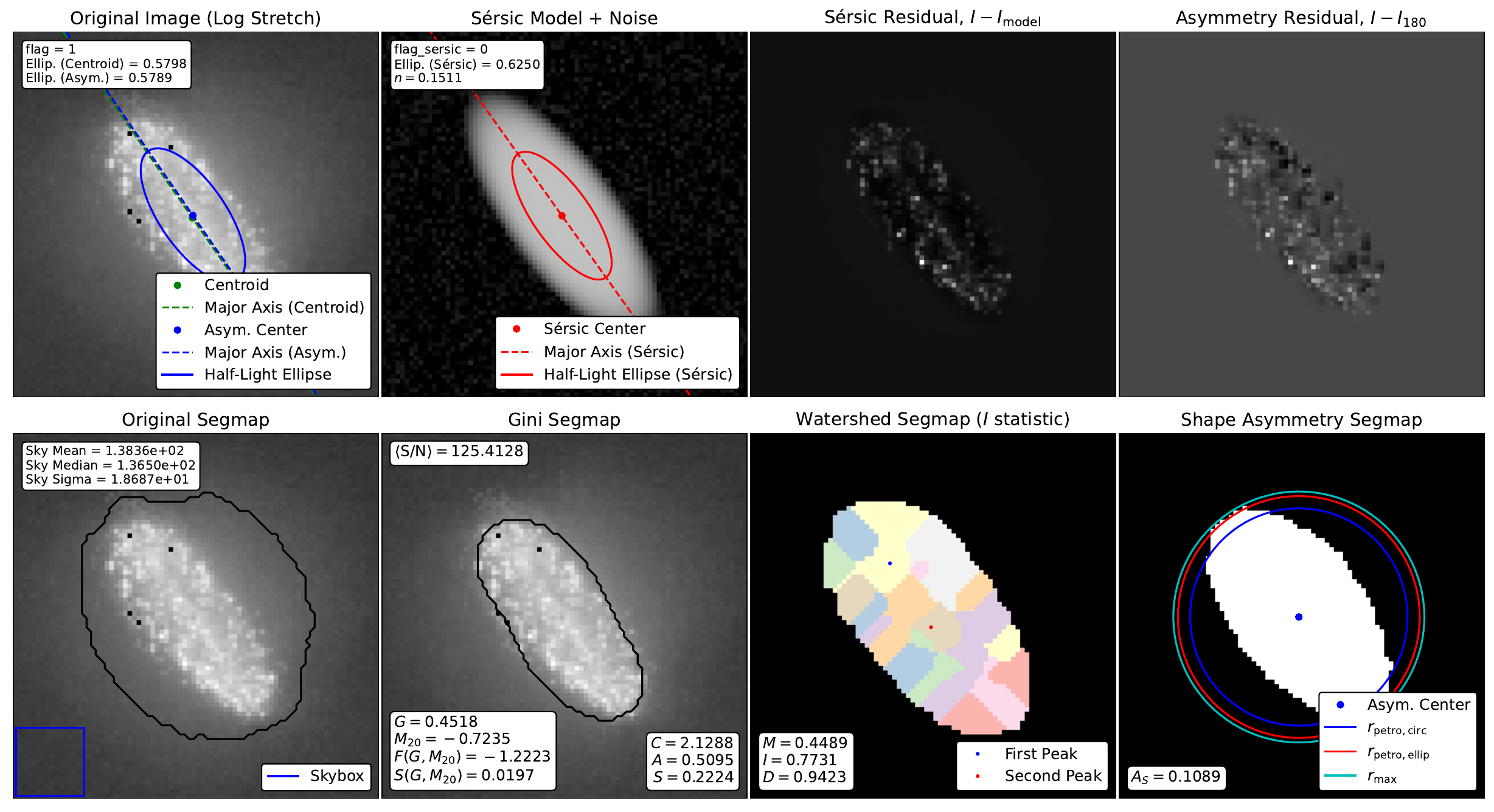}
\figcaption{Our segmentation and fitting of the mock images to extract morphological properties using \texttt{StatMorph} on the same example galaxy and filter as in Figure~\ref{fig:matching-mock-images}.}
\label{fig:plot-of-statmorph-properties}
\end{figure*}

The concentration parameter ($C$) also measures the concentration of light in a galaxy, although it is relative to the center of the galaxy. A higher $C$ is more likely to be an early-type galaxy, with a higher proportion light centralized. It is measured as

\begin{equation*}
    C = 5\log_{10}\left(\frac{r_{80}}{r_{20}}\right),
\end{equation*}

where $r_{80}$ and $r_{20}$ are the circular radii containing 80\% and 20\% of the total flux, respectively. The total flux for this measurement is defined as that within a radius of 1.5$r_p$ (Petrosian radii). This masking is performed for all CAS parameters.

The asymmetry parameter (A) measures the rotational symmetry of the galaxy. Higher values of $A$ indicate disturbed morphology or tidal tails, and lower values indicate settled morphology on large scales. It is calculated as

\begin{equation*}
    A = \frac{\sum_{i,j} |I(i,j)-I_{180}(i,j)|}{\sum_{i,j} |I(i,j)|} - A_B,
\end{equation*}

for all pixels ($i$,$j$) within a radius of $1.5r_p$, where $I(i,j)$ is the image flux at pixel ($i$,$j$), $I_{180}(i,j)$ is the image rotated by 180$^{\circ}$, $A_B$ is the background asymmetry. The asymmetry parameter can be used to define a galaxy merger, where $A > 0.35$ (more asymmetric galaxies) are identified as galaxy mergers \citep{Conselice_2003}.  

The smoothness parameter ($S$; also called clumpiness) measures the fraction of light found in clumpy distributions. It is thus a measure of recent star formation. It is limited by resolution and thus not well applied to distant, poorly-resolved galaxies. It is calculated as

\begin{equation*}
    S = \frac{\sum_{ij} |I(i,j)-I_s(i,j)|}{\sum_{i,j}|I(i,j)|} - S_B,
\end{equation*}

for all pixels ($i$,$j$) within a radius of $1.5r_p$, where $I_s(i,j)$ is the smoothed image obtained from a boxcar kernel of size 0.25$r_p$ \citep{Lotz_2008b}, and $S_B$ is the background smoothness. We limit the boxcar width to a minimum of 2 pixels to ensure that there is smoothing even for small or poorly-resolved galaxies. Since central regions of galaxies are typically highly concentrated, pixels in a central circle of radius 0.25$r_p$ are excluded from the sum. 

The shape asymmetry ($A_S$) is measured in the same way as the asymmetry parameter, but it is measured with a binary detection mask. This method removes relative flux brightness from the calculation, giving a measurement of asymmetry purely from the morphology rather than the light distribution. The shape asymmetry can measure fainter tidal features when compared to $A$. Like $A$, the shape asymmetry parameter can also be used to define a galaxy merger, where $A_S > 0.2$ are identified as galaxy mergers \citep{Pawlik_2016}.

The Sérsic index ($n$) is a parametric measurement which requires modelling the light profile of the galaxy. The surface brightness Sérsic profile of a galaxy is

\begin{equation*}
    I(R) = I_e \exp\left(-b_n\left[\left(\frac{R}{R_e}\right)^{\nicefrac{1}{n} - 1}\right]\right),
\end{equation*}

where $R_e$ is the half-light radius, $I_e$ is the surface brightness at that radius, and $b_n$ is a constant function of $n$. To obtain $n$, the Sérsic profile is fit to the galaxy light profile. Typically, values of $n < 2.5$ indicate spiral galaxies and values of $n > 2.5$ indicate elliptical galaxies. Since it assumes circular symmetry, $n$ tends to be less effective for irregular galaxies. 

Since different morphological statistics are better suited for different scenarios, no single morphological measurement is best for MBH merger host galaxy classification. Instead, a combination of non-parametric and parametric measurements can be used to account for a variety of types of galaxies and merger stages. We show an example of our morphological measurement extraction in Figure~\ref{fig:plot-of-statmorph-properties}, applied to the mock image of a galaxy in our sample. 

\begin{figure*}[ht]
\centering
    \includegraphics[width=0.305\textwidth, keepaspectratio]{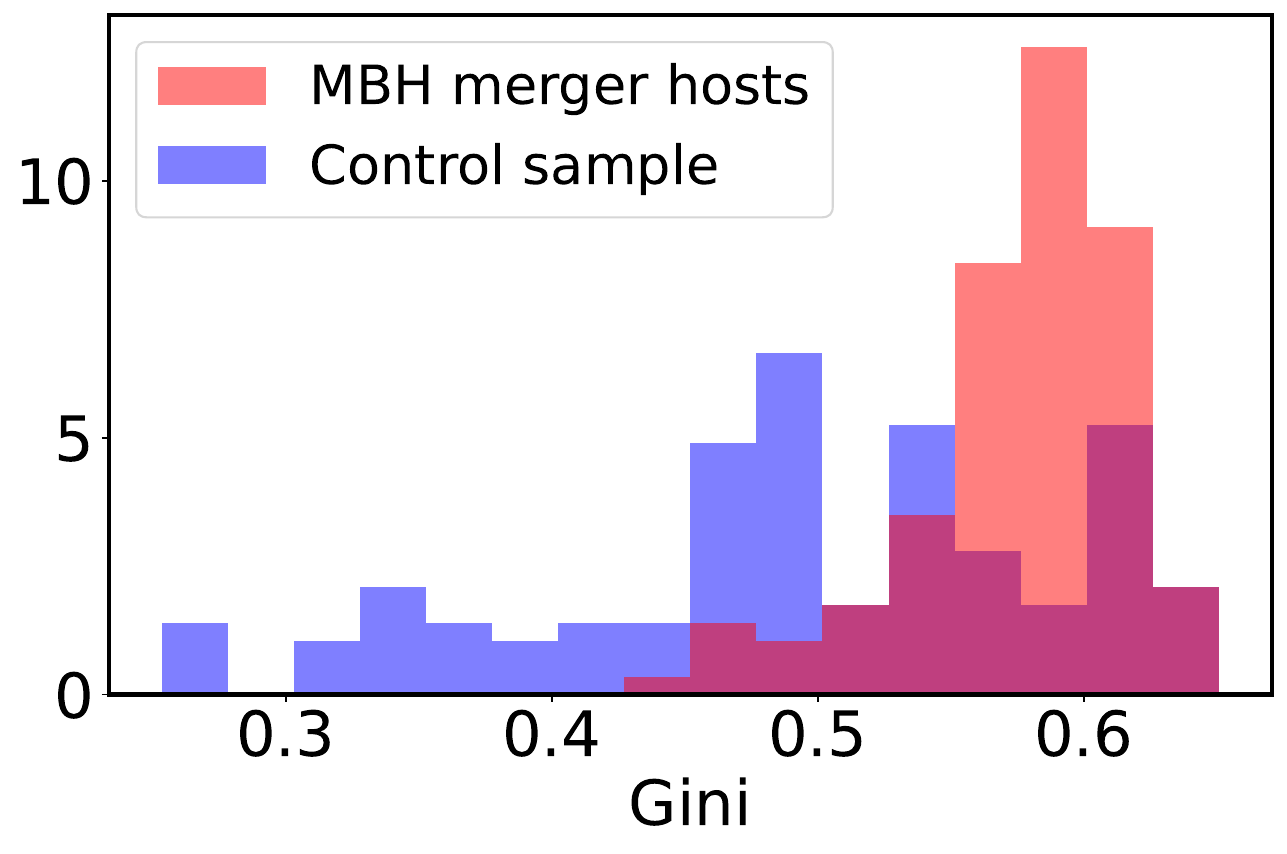} 
\hspace{5pt}
    \includegraphics[width=0.31\textwidth, keepaspectratio]{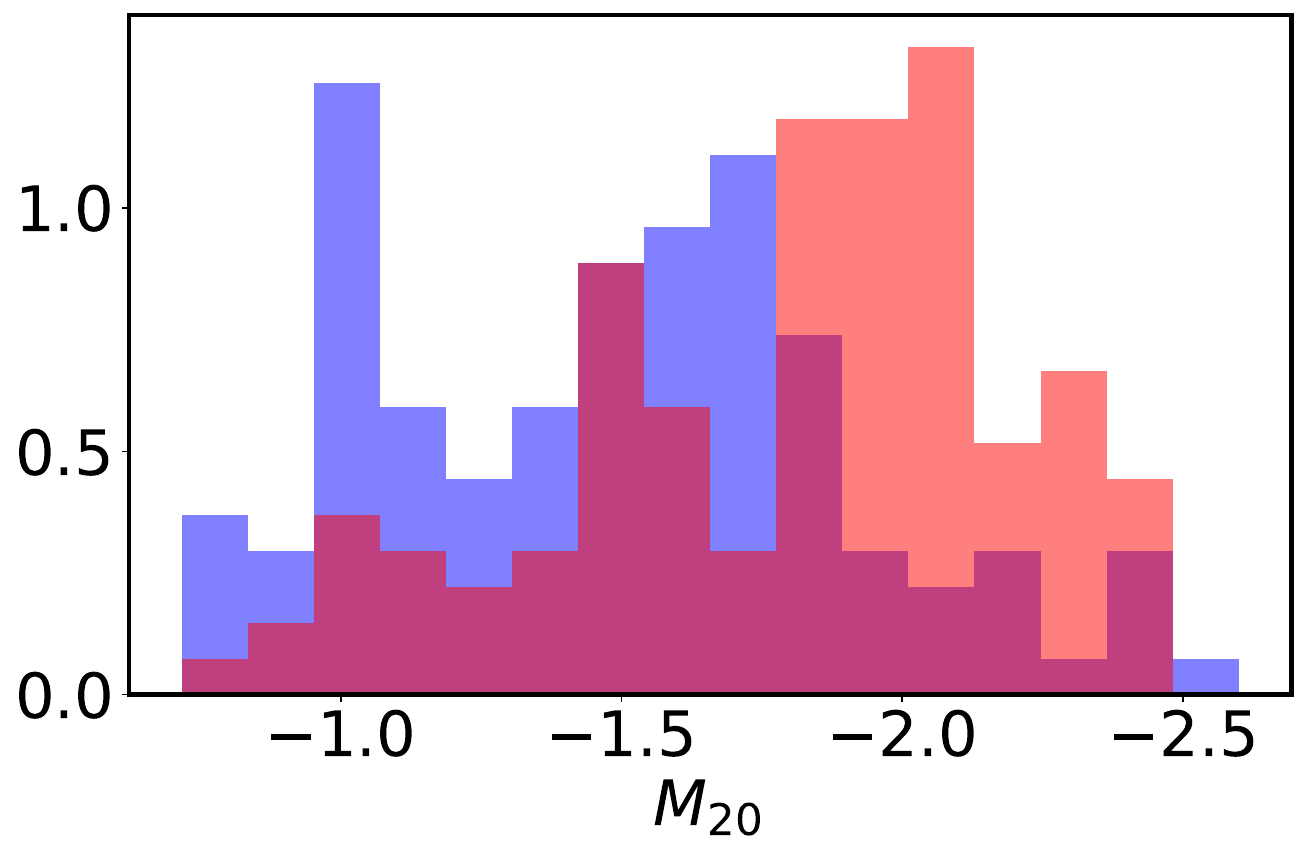}\\  
\vspace{10pt} 
    \includegraphics[width=0.315\textwidth, keepaspectratio]{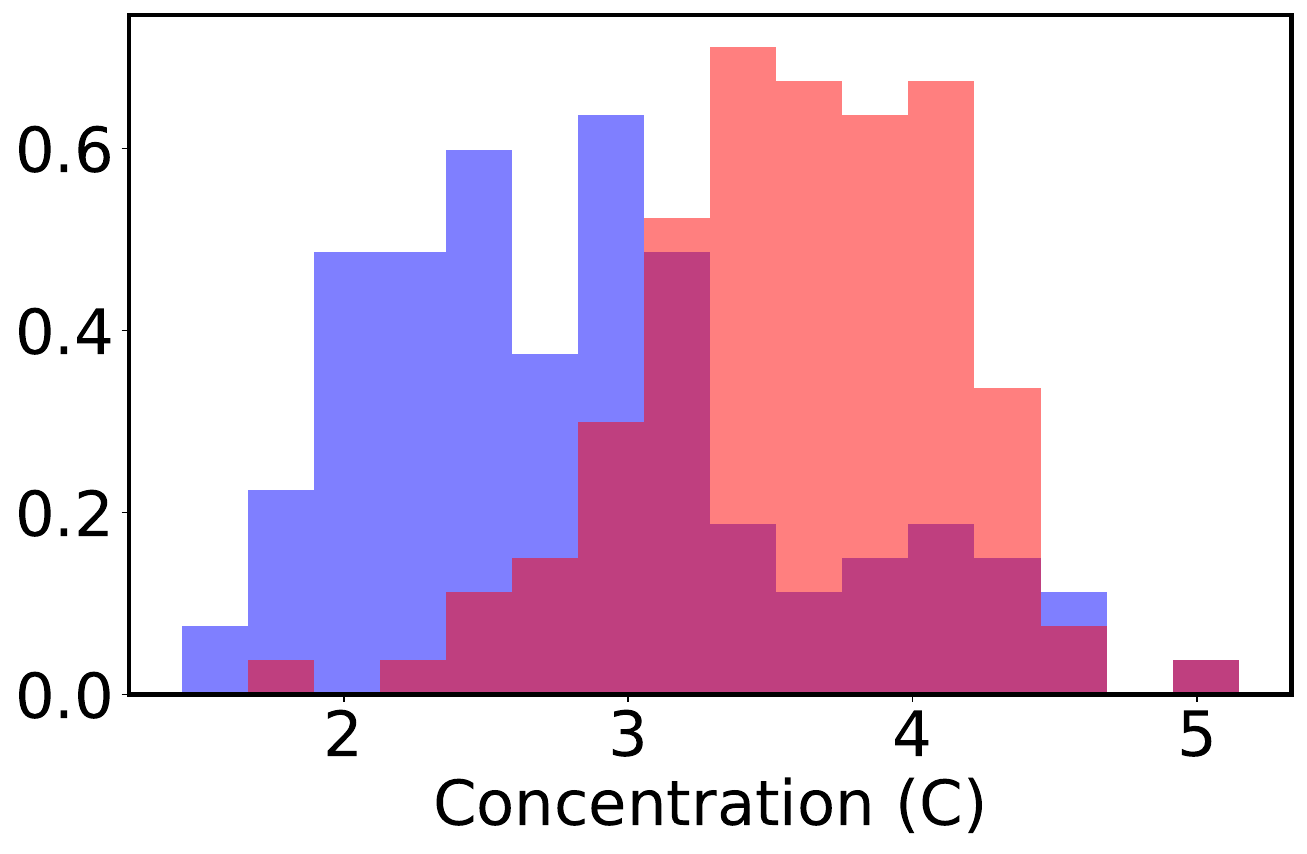} 
\hspace{10pt}
    \includegraphics[width=0.3\textwidth, keepaspectratio]{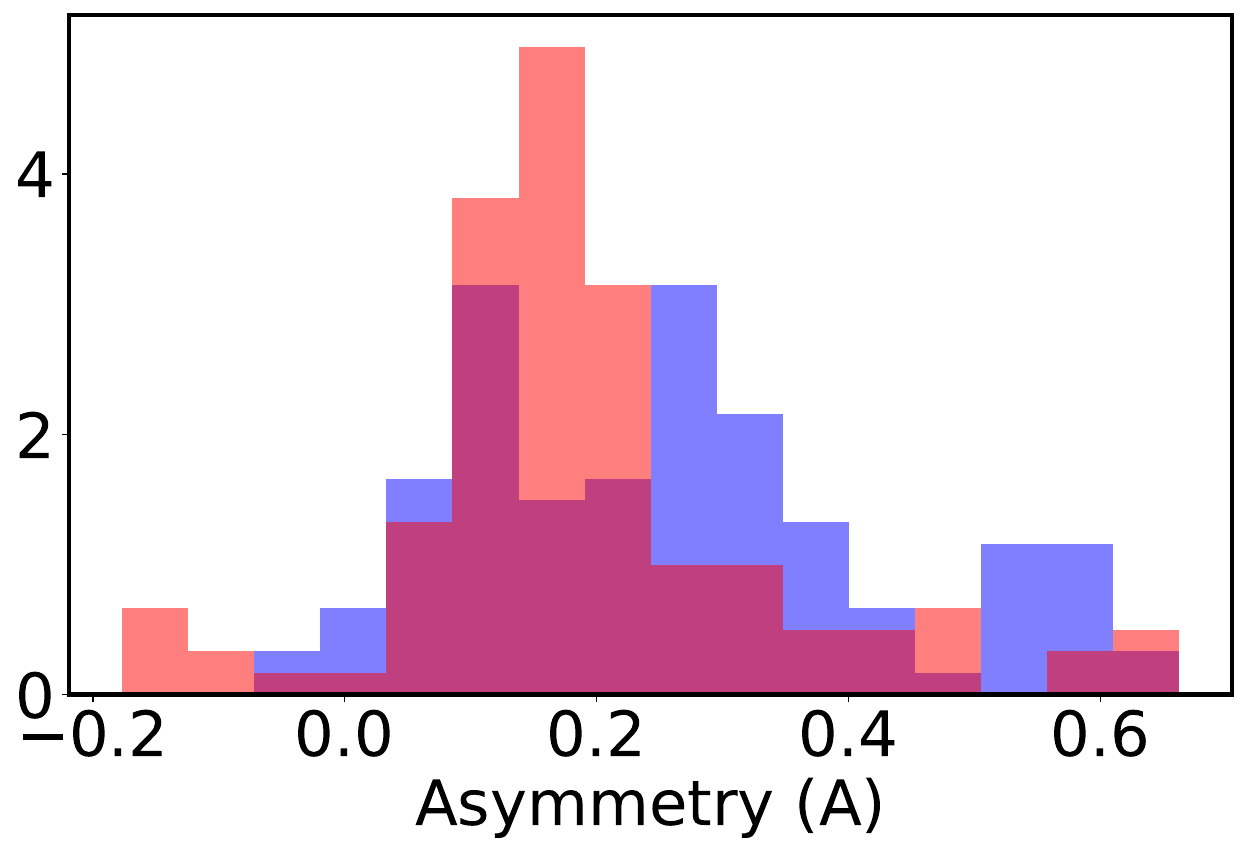} 
\hspace{5pt}
    \includegraphics[width=0.3265\textwidth, keepaspectratio]{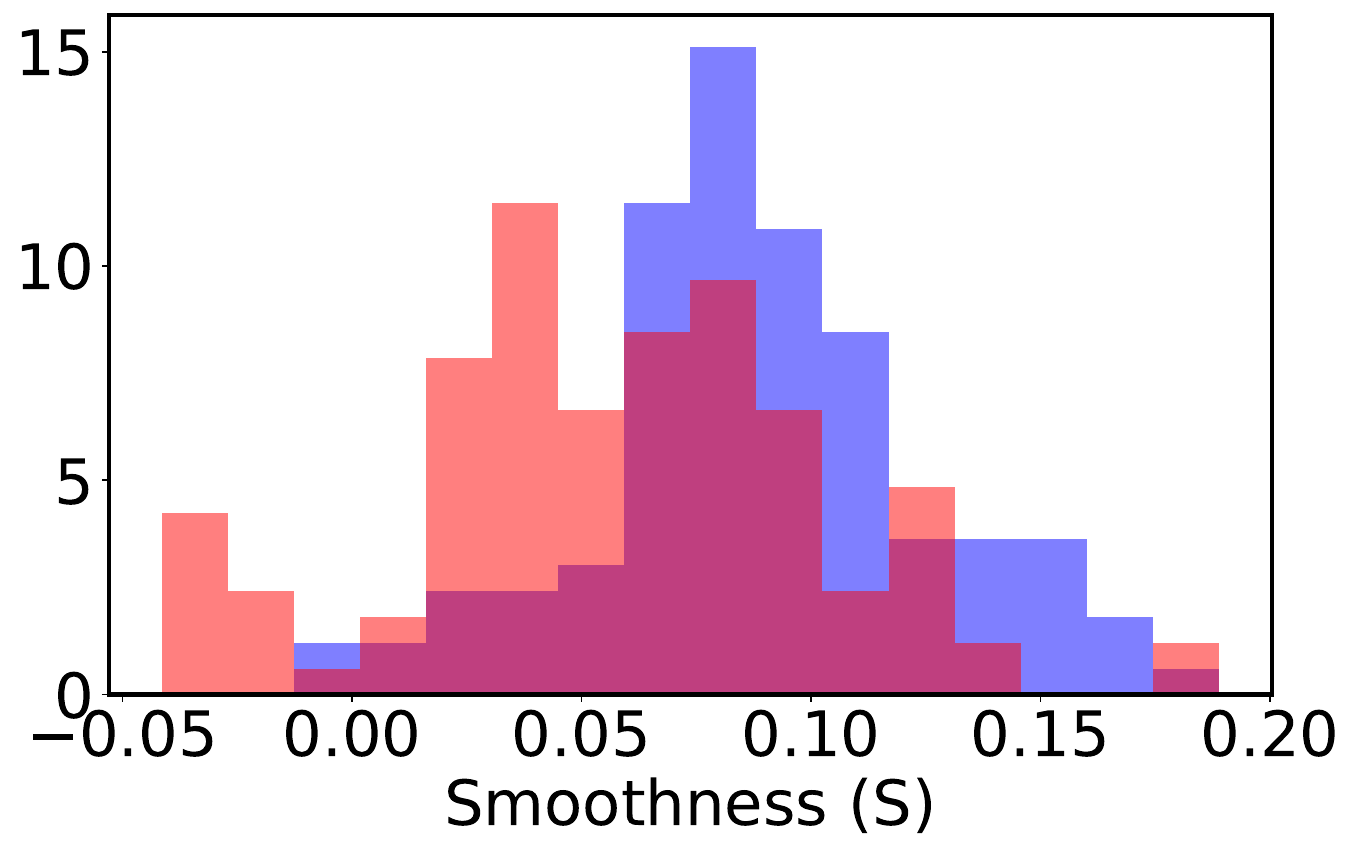} \\
\vspace{10pt} 
    \includegraphics[width=0.30\textwidth, keepaspectratio]{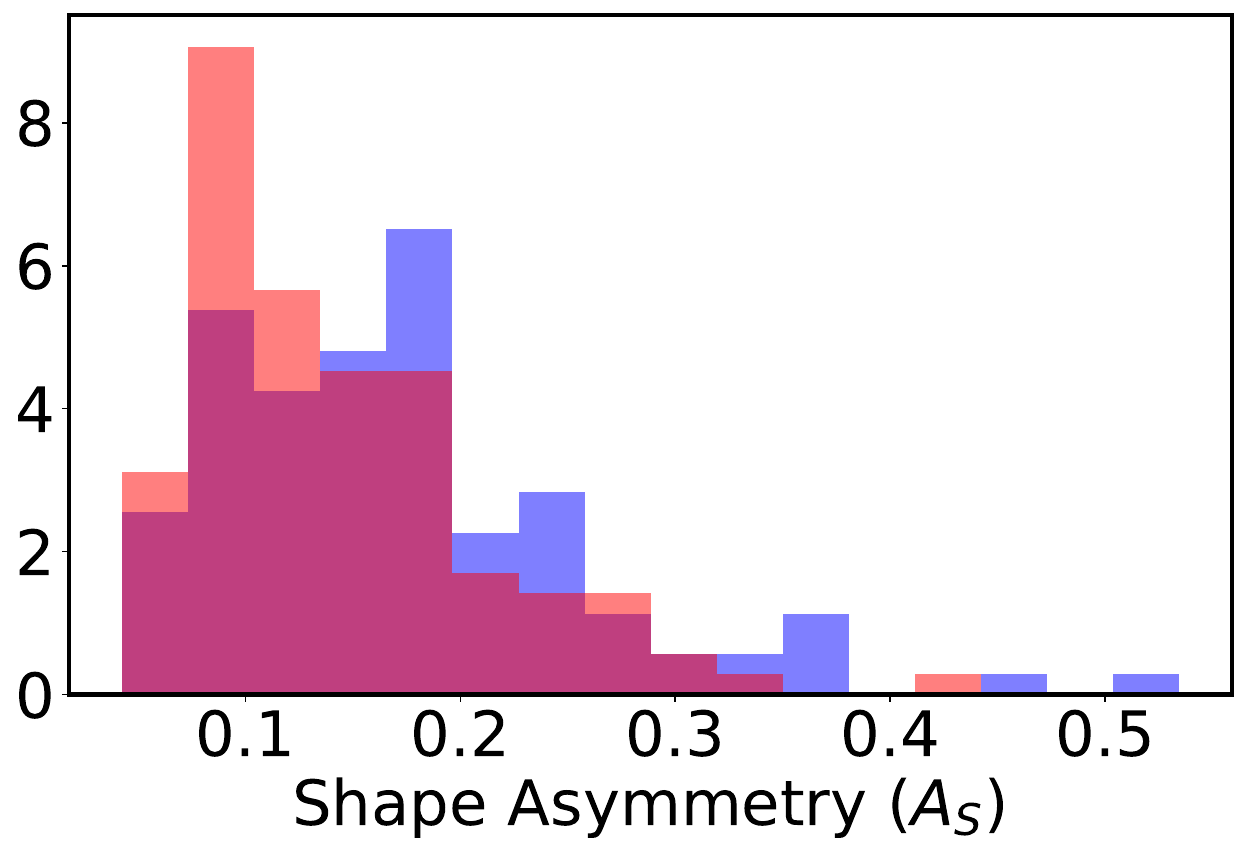} 
\hspace{10pt}
    \includegraphics[width=0.325\textwidth, keepaspectratio]{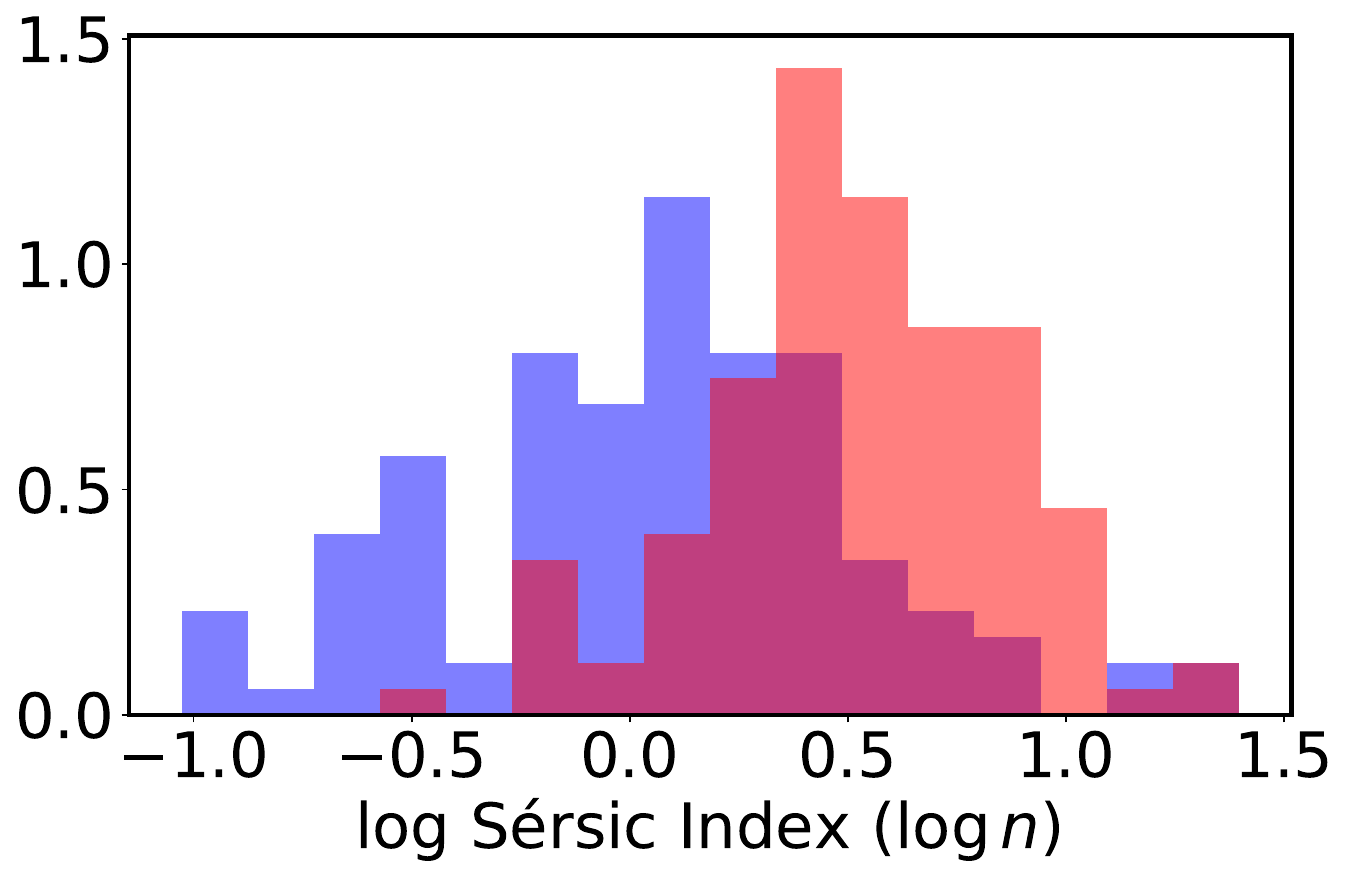} 
\figcaption{Normalized histograms of the seven morphological parameters from our $g$-band mock images of the subsample of MBH mergers host galaxies (red) with high mass ratios ($\nicefrac{M_2}{M_1} \geqslant 0.5$) and high chirp masses ($M_\text{chirp} \geqslant 10^{8.2} M_\odot$), in comparison to a corresponding mass- and redshift-matched control sample (blue). Although these histograms
show significant differences between the MBH merger host galaxies and the control sample, no single morphological statistic can discriminate between the two classes with high accuracy, thus motivating our use of LDA on combinations of these morphological measurements in Section~\ref{subsec:LDA}.}
\label{fig:meas-hists}
\end{figure*}

Figure \ref{fig:meas-hists} shows histograms of the morphological parameters for all $g$-band images of high mass ratio ($\nicefrac{M_2}{M_1} \geqslant 0.5$) and high chirp masses ($M_\text{chirp} \geqslant 10^{8.2} M_\odot$) MBH merger hosts, and a corresponding mass- and redshift-matched control sample. These histograms show significant differences between the MBH merger host galaxies and the control sample, although no single morphological statistic can differentiate between MBH merger host galaxies and control sample with high accuracy. Instead, we will apply LDA to these morphological measurements in Section \ref{subsec:LDA} to find the linear combination of these measurements that best discriminates between these two classes. Note that we use the Sérsic index in log scale ($\log n$) since its distribution appears Gaussian, which LDA is fairly robust to.

\begin{figure*}[ht]
\centering
    \includegraphics[width=0.66\textwidth, keepaspectratio]{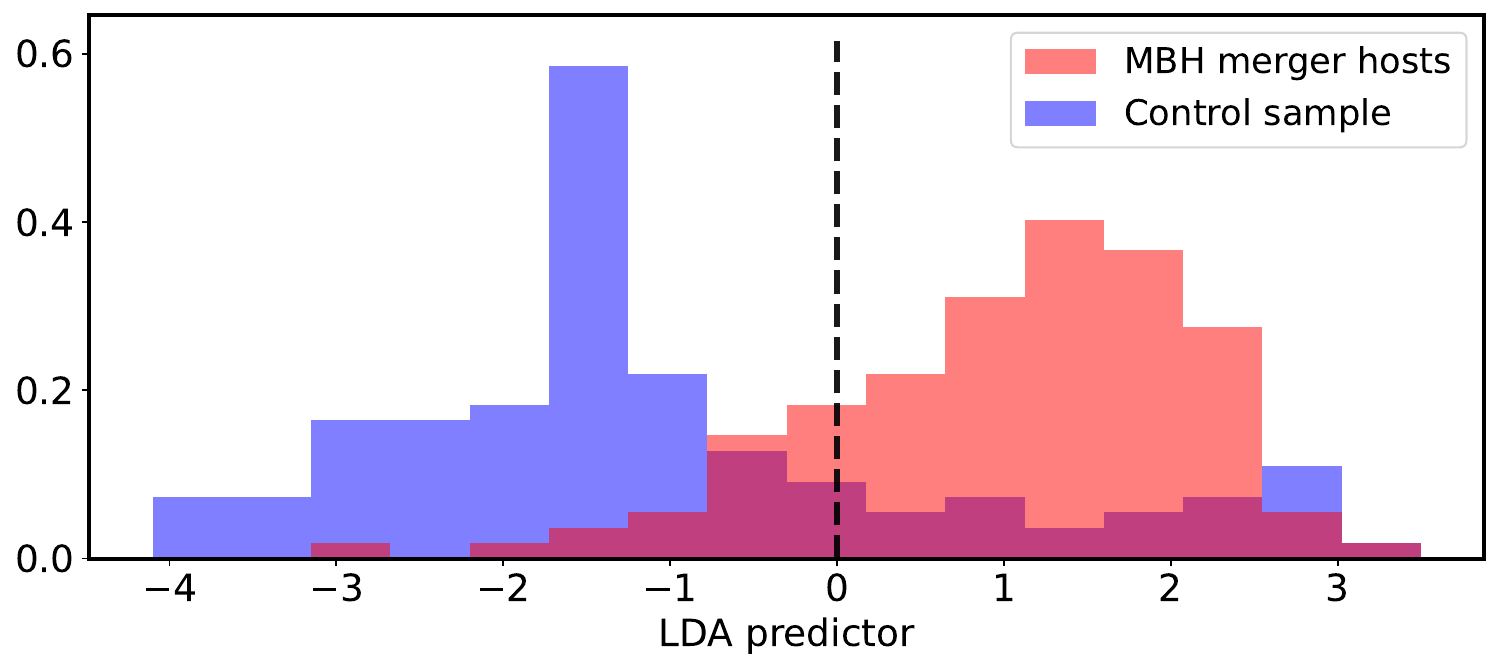} 
    \hspace{10pt}
    \includegraphics[width=0.30\textwidth, keepaspectratio]{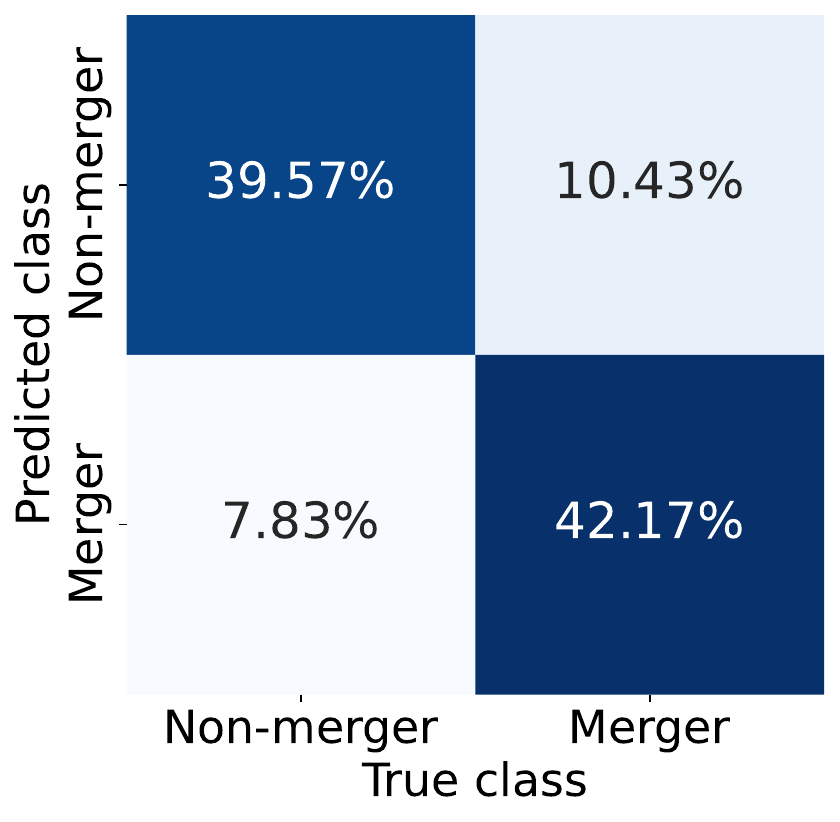}
\figcaption{Histogram of the LDA predictor (left) for same subsample of MBH merger hosts (red) and control subsample of non-merger hosts (blue) as shown in Figure~\ref{fig:meas-hists}. The LDA predictor maximally discriminates the MBH merger host sample from the control sample, with a decision boundary indicated by the dashed black line (with negative LDA predictor indicating a control sample galaxy, and positive LDA predictor indicating a MBH merger host). The confusion matrix (right) shows that the LDA predictor function classifies mergers in this subsample with an accuracy of 81.8\% and a precision of 84.4\%.}
\label{fig:ld1-example}
\end{figure*}

\subsection{Linear Discriminant Analysis}\label{subsec:LDA}

We use a linear discriminant analysis (LDA) on the classified samples to maximize the separation in morphological properties between our MBH merger host galaxies and control sample. The LDA algorithm finds the equation of a hyperplane that optimally separates the data in a hyperspace formed from the morphological measurements. The direction perpendicular to this hyperplane defines a linear function of the morphological measurements, which we can use to classify the galaxies into MBH merger hosts and control sample galaxies. The function is normalized such that the LDA denotes the signed distance of a data point to the hyperplane. Applying the function to the whole dataset allows us to create a confusion matrix containing the predicted class to true class statistics. The sum of the diagonal of this confusion matrix is the accuracy (the ratio of training data predicted by the LDA correctly). To perform the LDA, we use the \texttt{sklearn} Python package \citep[][]{scikit-learn}.

In addition to the seven `primary' morphological measurements (Gini, $M_{20}$, etc.), we also include `interaction terms' in the LDA, which are multiples of combinations of morphological measurements (e.g., $Gini \times M_{20}$). Since the predictive ability of the LDA is increased when there is an absence of multicollinearity, our inclusion of these interaction terms account for cross-correlations between the seven morphological measurements. These interaction terms add the combined effects of multiple measurements, and remove cross-correlation effects from the LDA coefficients \citep{James_2013}. This allows us to study the imaging predictors directly in Section \ref{subsec:coeffanalysis}.

Prior to being input into the LDA, the morphological parameters are whitened (i.e., normalized). To do this, we subtract the mean and divide by the standard deviation of the sample and control data for each data point. This prevents any one particular variable from dominating the LDA predictor, since the means and standard deviations of the morphological measurements can change drastically across the measurements. 

To obtain uncertainties on the LDA classification, we use repeated stratified $k$-fold cross validation. This method divides the sample of MBH merger host galaxies and control sample galaxies into $k$ bins of equal size. Stratification of the bins ensures that each fold of data set has the same proportion of each class label (MBH merger host or control), which is important to avoid class imbalances in the LDA. We use $k-1$ bins as input data to the LDA as training data, and the last bin as test data to test the resulting LDA. For each run of the LDA, the accuracy (fraction of correct classifications to the total number of classifications) and precision (the fraction of correct positive classifications) are determined as measures of the effectiveness of the LDA in classifying MBH merger host galaxies from the control sample. This process is then repeated a total of $n=10$ times with a different set of $k$ bins. We use the median across all runs, which provides a better representation of the LDA accuracy along with $1\sigma$ uncertainties. Through our tests, we find that $k=5$ gives a good balance between the LDA accuracy and uncertainties, and the LDA is robust to the assumed $k$ or $n$.

To reduce the number of parameters used in the LDA, we also use forward stepwise selection in addition to the $k$-fold cross validation. Since there are 28 morphological parameters used (including the interaction terms), we aim to reduce the number of parameters in the linear function returned from the LDA without significantly degrading its discriminating power. Since it is intractable to search all possible combinations of LDA parameters, we must choose a selection algorithm that quickly finds a suitable set of parameters. Through forward stepwise selection, we first begin with a model without any morphological parameters, and introduce a single parameter at a time. We then choose the model that maximizes the total LDA cross validation accuracy (i.e., minimizes the number of misclassifications). We repeat this process iteratively, increasing the set of parameters by one for each iteration, until the addition of any more of the parameters will decrease the accuracy. We only consider adding parameters that have distributions that have significant differences between the MBH merger hosts and control sample. Specifically, we only consider morphological measurements that pass a two-sample Kolmogorov–Smirnov test with a $p$-value of $\leqslant 0.003$ ($3\sigma$), rejecting the hypothesis that the MBH merger host sample is drawn from the same distribution as the control galaxy sample. This avoids including obviously-uninformative parameters. Furthermore, we confirm that the results are not due to overfitting by confirming that randomly shuffling the morphological parameters between the merger and control samples yields an LDA accuracy that is consistent with $\sim$50\%.

In Figure~\ref{fig:ld1-example}, we show the resulting LDA predictor for a high mass ratio ($\nicefrac{M_2}{M_1} \geqslant 0.5$) with high chirp masses ($M_\text{chirp} \geqslant 10^{8.2} M_\odot$) MBH merger sample, and its associated confusion matrix. The classification accuracy of the LDA can be directly calculated from the LDA predictor values. From cross-validation, we find that the LDA discriminates between our sample MBH merger hosts and our control sample with a mean accuracy of 82.6 $\pm$ 3.3\% and a mean precision of 83.5 $\pm$ 6.1\%. The linear equation found from the LDA in Figure~\ref{fig:ld1-example} is given by:

\begin{equation}\label{eq:LD1selection}
    \begin{split}
    \text{LDA predictor} = 1.23 Gini + 0.51 M_{20} + 0.52 C \\
    - 1.04 S -0.01.
    \end{split}
\end{equation}

This function represents the median-scoring LDA from the cross validation. We also compute the mean statistics for the cross validation, giving us a measure of uncertainty on the coefficients. For this subsample, we obtain mean and 1$\sigma$ uncertainties of $1.26 \pm 0.13$, $0.55 \pm 0.12$, $0.58 \pm 0.18$, and $-0.98 \pm 0.11$, for the Gini, $M_{20}$, Concentration ($C$), and Smoothness ($S$), respectively. We discuss these results in the following section.

\begin{figure*}[ht]
\centering
    \includegraphics[width=0.48\linewidth]{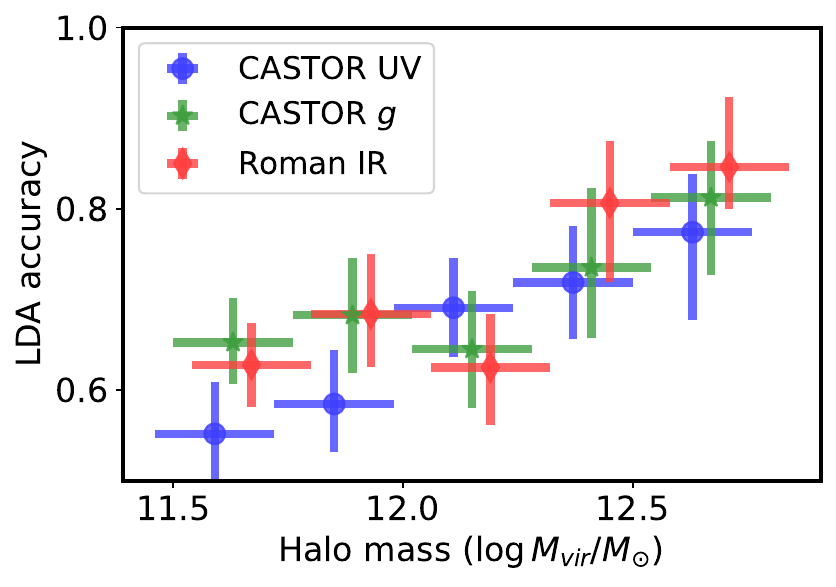}
    \hspace{10pt}
    \includegraphics[width=0.48\linewidth]{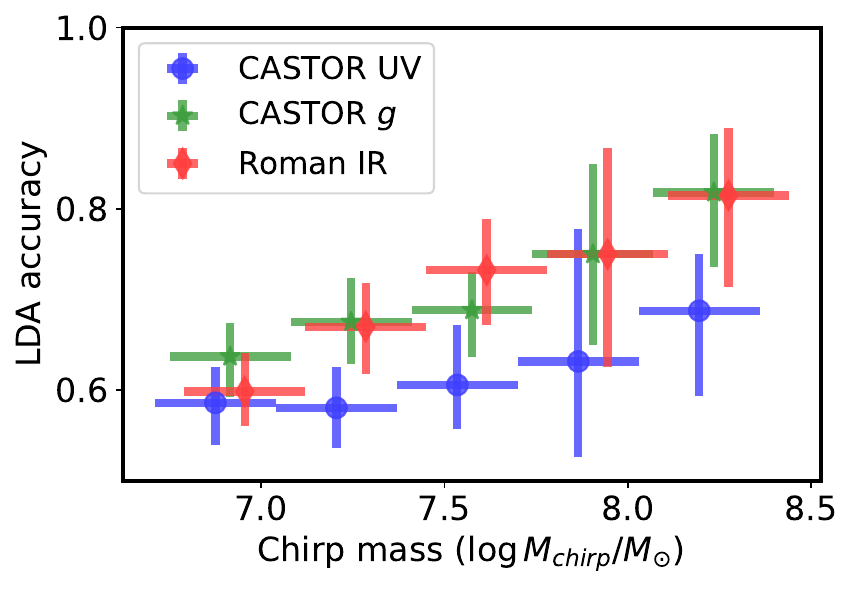} \\
    \vspace{2pt}
    \includegraphics[width=0.48\linewidth]{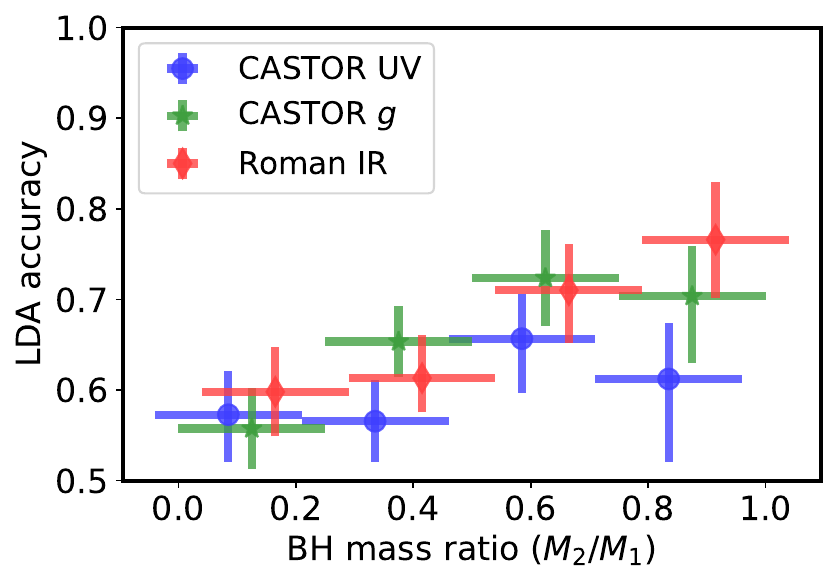}
    \hspace{10pt}
    \includegraphics[width=0.48\linewidth]{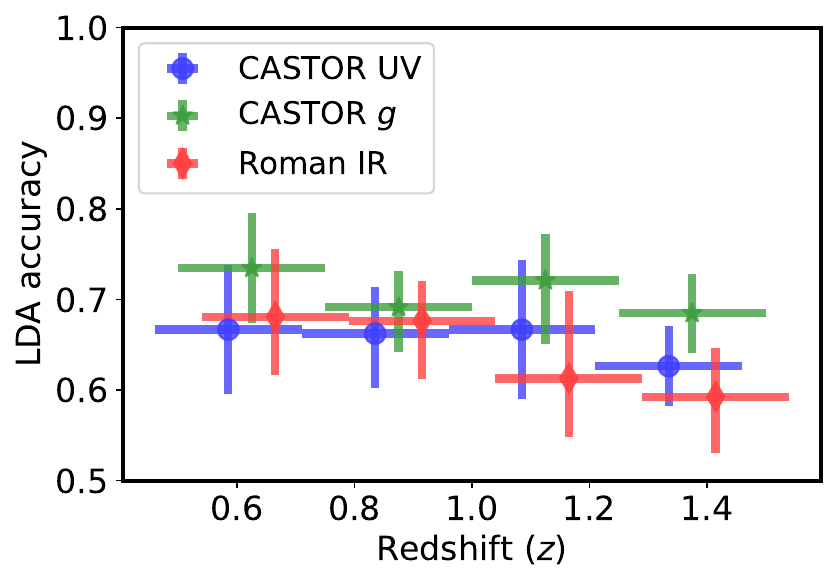}
\figcaption{The accuracy of the LDA predictor (ratio of correctly classified galaxies) as a function of halo mass (upper left), chirp mass (upper right), mass ratio (lower left), and redshift (lower right), separated by filter. The halo mass, chirp mass, and redshift subsamples only contain high mass ratio ($\nicefrac{M_2}{M_1} \geqslant 0.5$) mergers, and the MBH mass ratio only contains MBH chirp masses of $M_\text{chirp} \geqslant 10^{7.5} M_\odot$. The points indicate a median and the error bars represent the 1$\sigma$ spread. An LDA accuracy of 50\% would mean that the LDA predictive ability is no better than chance, and 100\% would mean that the LDA predicts the correct label with full accuracy. The accuracy of our morphology-based approach increases as a function of halo mass, chirp mass, and mass ratio, and weakly decreases with redshift.}
\label{fig:ldascore-vs-cuts}
\end{figure*}

\section{Results and Discussion} 
\label{sec:discussion}

\subsection{LDA Coefficient Analysis}\label{subsec:coeffanalysis}

A key advantage of using linear discriminant analysis (LDA) is that the returned coefficients provide direct insights on the strength of each parameter in discriminating between classes. Since we whiten our data, the coefficients can be interpreted directly as the relative importance between the measurements. Thus, from Eq.~\ref{eq:LD1selection}, we can see that the most important parameters are Gini, $M_{20}$, $C$, and $S$. This is visually consistent with Figure~\ref{fig:meas-hists}, where these morphological measurements best distinguish between the MBH merger hosts sample and the control sample. The sign of the coefficient is also important, because it indicates whether an increase in the parameter will increase (in the case of a positive) or decrease (in the case of a negative) the probability of the galaxy being identified as a MBH merger host when controlling for the other measurements. We discuss the coefficients from our subsample of high mass ratio and high chirp mass MBH mergers (from Figure \ref{fig:ld1-example}) in this section.

The Gini coefficient is positive and significantly different from zero (to $>$$3\sigma$). This implies that galaxies with very bright single or dual nuclei are more likely to be MBH mergers. This is both expected and consistent with the data in Figure~\ref{fig:meas-hists}.

Since $M_{20}$ and concentration ($C$) are generally negatively correlated, we must consider them together. The $M_{20}$ and $C$ coefficients are both positive and significantly different from zero. This implies that, when controlled for $M_{20}$, galaxies are more likely to be MBH merger hosts if they are early-type galaxies with high $C$, which is clear in Figure~\ref{fig:meas-hists}. When controlled for $C$, galaxies are more likely to be MBH merger hosts if they have higher $M_{20}$. This is not obvious from Figure~\ref{fig:meas-hists}, which does not control for $C$, but is expected because our MBH merger hosts subsample has less correlation between $M_{20}$ and $C$ than the matched control sample. 

The smoothness ($S$) coefficient is negative and significantly different from zero. This indicates that less clumpy galaxies, when controlled for the other parameters, are more likely to host MBH mergers. This can be visually confirmed in Figure~\ref{fig:meas-hists}.

While the log Sérsic index ($\log n$) of the the MBH merger host and control sample has a visually separate distribution in Figure~\ref{fig:meas-hists}, it does not appear in the LDA predictor equation. This is because $C$ and $\log n$ are correlated in our subsample, and thus only one is necessary to provide the same LDA accuracy. 

\subsection{Trends in Discriminating Power}\label{subsec:trends}

In this section, we investigate the efficacy of our morphology-based approach in discriminating between MBH merger hosts and control sample as a function of halo mass, chirp mass, mass ratio, redshift, and telescope filter. We track the efficacy using the LDA accuracy, which is the the percentage of correctly classified galaxies. For example, Equation~\ref{eq:LD1selection} has an LDA accuracy of $\gtrsim$80\%, which means that $\gtrsim$4 of every 5 of such galaxies are correctly identified using this method. We discuss the trends in LDA accuracy here, and interpret these results in greater detail in Section~\ref{subsec:interpretation}.

Figure~\ref{fig:ldascore-vs-cuts} shows that the accuracy of our morphology-based approach increases as a function of halo mass, chirp mass, and mass ratio. To compute the accuracy for the subsample in each bin in halo mass, chirp mass, and mass ratio, we also create a corresponding mass- and redshift-matched control subsample. For visual clarity, we only consider high mass ratio ($\nicefrac{M_2}{M_1} \geqslant 0.5$) MBH mergers for the halo mass, chirp mass, and redshift plots, while we only consider high MBH chirp masses ($M_\text{chirp} \geqslant 10^{7.5} M_\odot$) for the MBH mass ratio plot. 

Figure~\ref{fig:ldascore-vs-cuts} also shows that the LDA accuracy displays a weak negative trend with redshift. This may be in part due to the decrease in galaxy angular size as a function of redshift, causing galaxies to be less resolved in the images at higher redshifts, thus leading to worse LDA accuracies.

Finally, Figure~\ref{fig:ldascore-vs-cuts} shows that in general, UV imaging has slightly but systematically worse LDA accuracies than in \textit{g}-band and IR bands. We note that our mock images are generated in the galaxy rest-frame, and we do not include cosmological redshifting of the spectrum. Imaging at optical or IR wavelengths may extend the effectiveness of this method in lower chirp masses and mass ratios. 

\subsection{Dependence of Results on Time-Delay}\label{subsec:timescales}

Romulus25's spatial resolution numerically merges MBHs at separations of $\lesssim$700~pc, which is much larger than the separations at which gravitational waves will be detected by PTAs and \emph{LISA}. This introduces a time-delay between when the simulation designates the MBHs as merged (the `numerical merger') and when gravitational waves would be detected (the `physical merger'). This time-delay is highly uncertain and varies by merger, and can be $\sim$Gyrs. We thus investigate the dependence of our results as a function of time-delay.

To understand how the MBH merger signatures are affected as a function of the time since numerical merger, we reperform our analysis assuming a range of time-delays. Since time-delays below the resolution scale of our simulation are highly uncertain, we do not attempt to carefully calculate theoretical time-delays. Instead, we remain agnostic to the exact time-delay for each merger, and instead investigate how our results change as a function of the assumed time-delay. Specifically, we consider all snapshots of a host galaxy after its MBH has numerically merged, up to $\sim$1~Gyr. 

\begin{figure}[tb]
\begin{center}
    \includegraphics[width=0.48\textwidth, keepaspectratio]{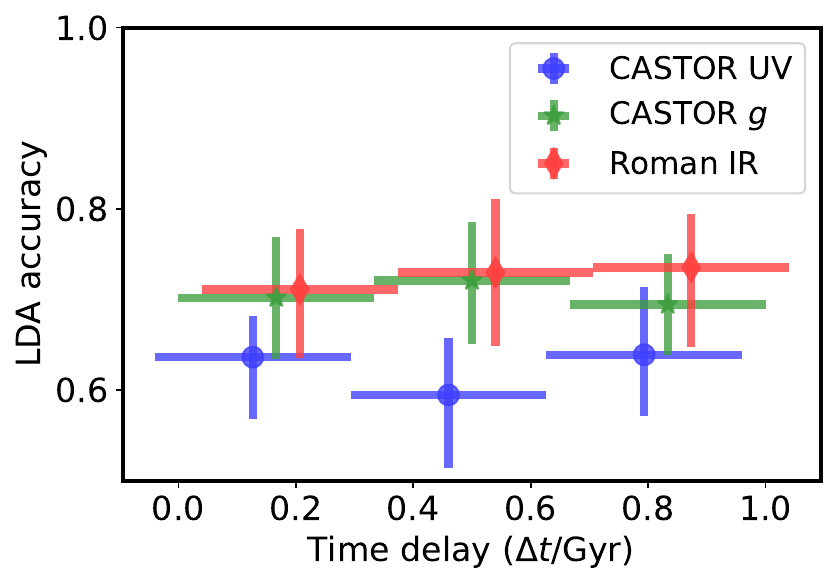}
\end{center}
\figcaption{The effect of the time-delay since numerical merger on the LDA predictor. This shows the accuracy of a representative (median score) LDA predictor as a function of the time delay for a high mass ratio ($\nicefrac{M_2}{M_1} \geqslant 0.5$) high chirp mass ($M_{chirp} > 10^{7.5} M_\odot$) merger subsample and corresponding control sample. This shows that galaxy morphological evidence for MBH mergers persists for at least $\sim$1~Gyr, longer than the expected delay-time to physical merger, and thus the accuracy of our morphology-based approach is likely to remain high at the time of gravitational wave detection.}
\label{fig:lda-vs-deltat}
\end{figure}

In Figure~\ref{fig:lda-vs-deltat}, we show the LDA accuracy as a function of time-delay after numerical merger. We consider the subsample of high mass ratio ($\nicefrac{M_2}{M_1} \geqslant 0.5$), high chirp mass ($M_\text{chirp} \geqslant 10^{7.5} M_\odot$) mergers as for Equation~\ref{eq:LD1selection}, which ensures a sufficiently large merger sample for this test. The LDA is performed separately for each time-delay bin, thus optimizing it for that specific time-delay range. Figure~\ref{fig:lda-vs-deltat} shows that galaxy morphological evidence for MBH mergers persist for at least $\sim$1~Gyr, significantly longer than reported in previous studies. We emphasize that the exact time-delay between numerical and physical merger for each MBH binary is highly uncertain, and can be between 0.1 to 10~Gyrs \citep{Volonteri_2020, Li_2022}. In Figure~\ref{fig:lda-vs-deltat}, we have chosen to test the accuracy of our morphology-based approach only up to time-delays of 1~Gyr, which saves on the computational costs of performing our radiative transfer simulations while still enabling a direct comparison to previous results. We discuss the possible reasons why our LDA accuracy remains high for time-delays of at least 1~Gyr, and compare to previous results in Section~\ref{subsec:interpretation} below.

\subsection{Interpretation and Comparison to Previous Work}\label{subsec:interpretation}

We suggest that the most distinct morphological characteristics of MBH merger host galaxies at high chirp masses and mass ratios are permanent features (such as the presence of a galaxy bulge), while galaxy merger-induced morphological disturbances play little to no role in the LDA. Our results in Section~\ref{subsec:timescales} suggest that the morphological signatures of MBH merger host galaxies persist for at least $\sim$1~Gyr after the numerical merger. This may at first glance appear to be in tension with previous studies, which instead suggest that these signatures will disappear on timescales of less than a few hundred Myrs after the numerical merger, well before the physical MBH merger when gravitational waves would be detected \citep{Volonteri_2020, DeGraf_2021}. However, those studies specifically studied the use of signatures of galaxy mergers for identifying MBH merger host galaxies. Specifically, \citet{DeGraf_2021} compares the morphologies of MBH merger host galaxies to a control sample using using the empirical galaxy merger discriminant $S(Gini, M_{20})$ from \citet{Lotz_2008b}, defined as

\begin{equation*}
    S(Gini, M_{20}) = 0.139 M_{20} + 0.990 Gini - 0.327,
\end{equation*}

where merging galaxies generally have $S(Gini, M_{20})>0$ and non-merging galaxies have $S(Gini, M_{20})<0$. Similarly, \citet{Volonteri_2020} visually search for disturbed morphologies in MBH merger host galaxies in their mock images, after more carefully taking into account the delay-time to physical merger. In contrast, our approach here is agnostic to the connection between MBH mergers and galaxy mergers. Our LDA predictor is thus also sensitive to more permanent morphological differences between MBH merger host galaxies and the control sample. If the unique morphological signatures of MBH merger host galaxies fade over time (such as disturbed morphologies from the preceding galaxy mergers), the LDA accuracy would decrease as a function of time-delay such that MBH host galaxies become indistinguishable from the control galaxy sample. The relatively stable LDA accuracies as a function of time-delay that we find in Figure~\ref{fig:lda-vs-deltat} is thus evidence that the LDA is actually identifying permanent morphological differences.

\begin{figure}[tb]
\begin{center}
    \includegraphics[width=0.47\textwidth, keepaspectratio]{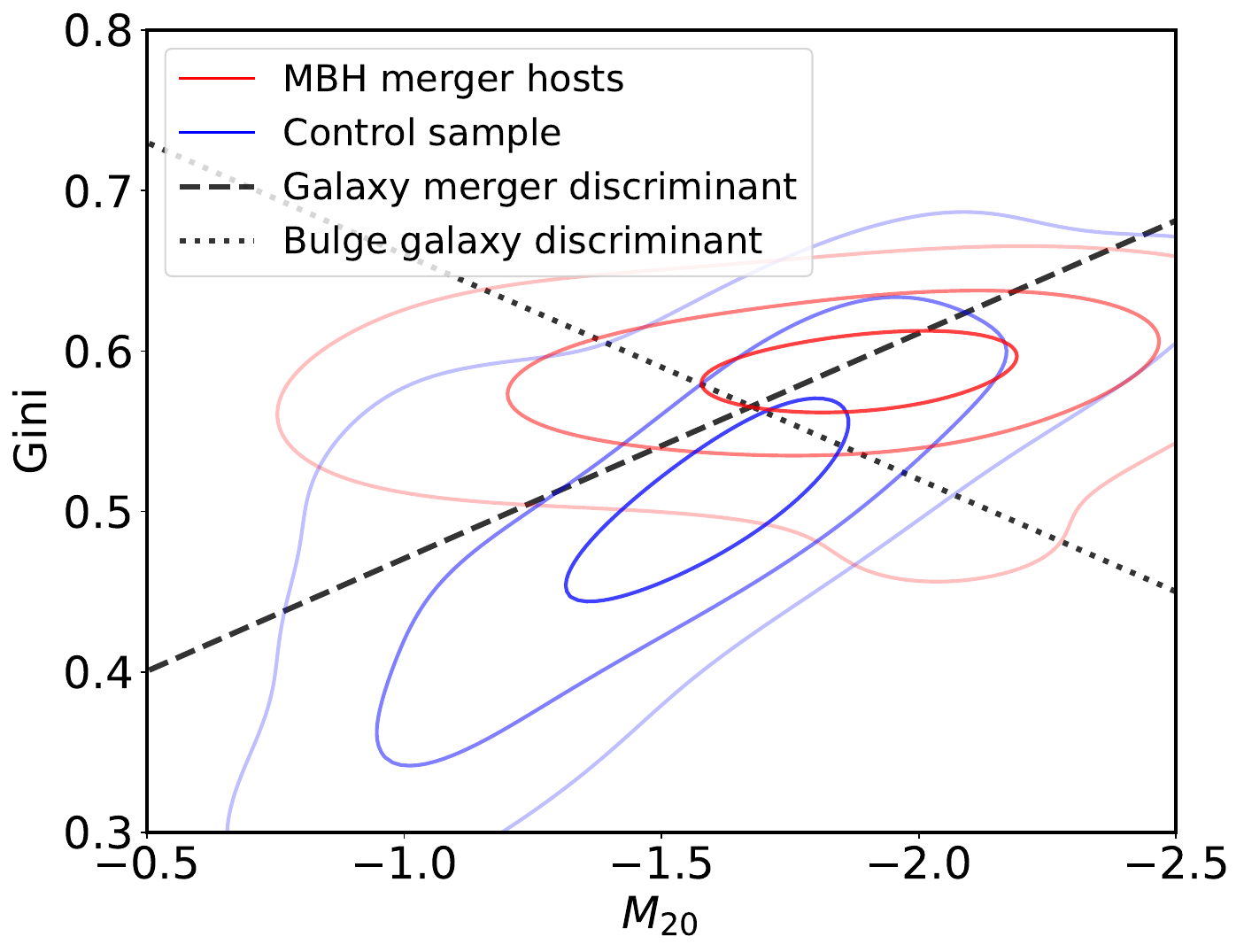}
\end{center}
\figcaption{The $M_{20}$--$Gini$ relation for our samples. To best demonstrate the difference, we only consider a subsample of high mass ratio MBH mergers ($\nicefrac{M_2}{M_1} \geqslant 0.5$) with high chirp masses ($M_\text{chirp} \geqslant 10^{8.2} M_\odot$) (red contours), with a corresponding mass- and redshift-matched control (blue contours). The contours enclose 15.9\% ($-1$$\sigma$), 50\% (median) and 84.1\% (1$\sigma$) of the MBH merger host and control sample. The dashed black line indicates the empirical galaxy merger discriminant line ($S(Gini, M_{20}) = 0$), above which would be classified as a galaxy merger. The dotted black line indicates the empirical bulge discriminant line ($F(Gini, M_{20}) = 0$), above which would be classified as having a bulge. The bulge discriminant line's clear ability to separate the MBH merger host sample from the control sample suggests that the presence of bulges may play a dominant role as morphological signatures of MBH merger host galaxies in the LDA, while galaxy merger signatures have little to no role.
\label{fig:gini-m20-relation}}
\end{figure}
 
The permanent morphological differences between the MBH merger host galaxies and control sample are likely to be related to galaxy bulges. Figure~\ref{fig:gini-m20-relation} compares Gini and $M_{20}$ of our MBH merger host galaxies sample to that of the control sample, for the subsample with high chirp mass and mass ratio. We also show the empirical merger discriminant line $S(Gini, M_{20}) = 0$, where galaxies lying above this line would be classified as mergers, while those below would not. Our MBH merger hosts sample lies slightly but systematically above this merger discriminant line, similar to the finding of \citet[][see their Figure~9]{DeGraf_2021}. However, we have already shown in Figure~\ref{fig:lda-vs-deltat} that the morphological differences between MBH merger host galaxies and the control sample identified by the LDA do not change over time, and thus these differences cannot be morphological disturbances from galaxy mergers. Instead, we also show in Figure~\ref{fig:gini-m20-relation} the empirical galaxy bulge discriminant $F(Gini, M_{20})$ from \citet{Lotz_2008b}, defined as

\begin{equation*}
    F(Gini, M_{20}) = -0.693 M_{20} + 4.95 Gini - 3.96.
\end{equation*}

Similar to the galaxy merger discriminant line, galaxies lying above the galaxy bulge discriminant line ($F(Gini, M_{20}) > 0$) have prominent bulges, and those below do not. Figure~\ref{fig:gini-m20-relation} shows that the bulge discriminant clearly separates MBH merger host galaxies from the control sample, thus implying that the permanent morphological differences between these two samples are related to the presence of bulges.

Our interpretation also explains the LDA accuracy trends in Figure~\ref{fig:ldascore-vs-cuts}, which show that the LDA accuracy increases as a function of halo mass, chirp mass and mass ratio, and decrease as a function of redshift. These trends can be explain by a scenario in which major (i.e., nearly equal-mass) mergers of massive galaxies form prominent classical bulges, leading to the formation of MBH binaries with high mass ratios, and eventually MBH mergers with high chirp masses. Thus, for MBH binaries and mergers with high chirp masses and mass ratios, the host galaxy bulges are prominent and serve as the key morphological signature, causing the LDA accuracy to be high. In contrast, MBH binaries and mergers with low mass ratios result from minor galaxy mergers that do not produce prominent classical bulges, thus leading to low LDA accuracies. Similarly, MBH binaries and mergers with low chirp masses (and halo masses) result from mergers of lower-mass galaxies, for which classical bulges may be more difficult to detect for a variety for reasons, also causing the LDA accuracy to be low. For example, the smaller angular size of bulges in lower-mass galaxies makes them more difficult to resolve in imaging. Furthermore, in lower-mass galaxies, pseudobulges from secular processes are more common than classical bulges from galaxy mergers \citep[e.g.,][]{Fisher_2011}, which can confuse the LDA predictor. These reasons can also explain why the LDA accuracy decreases with redshift, since at higher redshifts, bulges are more difficult to resolve and lower-mass galaxies are more abundant.

Finally, a scenario in which the host galaxies of MBH mergers are more likely to have prominent classical bulges may also explain our findings in Figures~\ref{fig:ldascore-vs-cuts} and \ref{fig:lda-vs-deltat} that the LDA accuracies are systematically lower in UV imaging relative to optical and IR. If the distinct morphological features of MBH merger host galaxies are indeed classical bulges (as suggested by their high Sérsic indices in Figure~\ref{fig:meas-hists}), then UV imaging would be a poor tracer of the morphology of this old stellar population, leading to lower LDA accuracies.

In summary, the picture that emerges from our results is that major mergers of massive galaxies produce galaxies with prominent classical bulges that host MBH binaries with high mass ratios. The morphological disturbances from the galaxy merger disappear well before the physical MBH merger, when gravitational waves will be emitted with high chirp mass. For these MBH binaries and mergers with high mass ratios and chirp masses detected in gravitational waves, the prominent classical bulges in their host galaxies built during the preceding galaxy merger act as permanent signposts that can be used to identify the exact host galaxy to high accuracies. This picture is consistent with and extends previous work \citep{Volonteri_2020, DeGraf_2021}, and may not be surprising, as theoretical models of galaxy evolution have suggested that galaxies with more luminous bulges indeed have higher rates of MBH mergers \citep[e.g.,][]{Haehnelt_2002}. We defer a more detailed study of the properties of the bulges in our simulated galaxies for future work.

\subsection{Caveats}\label{subsec:caveats}

The high accuracies of $\gtrsim$80\% we achieve in identifying galaxies hosting merging MBHs through morphology are only applicable for relatively-massive (chirp mass of $\gtrsim$10$^{8.2}$ $M_\odot$) MBH binaries and mergers with high mass ratios ($\gtrsim$0.5). The accuracy of our approach is lower for MBH binaries and mergers that are detected in gravitational waves with lower chirp masses and lower mass ratios, as demonstrated in Figure~\ref{fig:ldascore-vs-cuts}. Thus, MBH binaries detected by PTAs and particularly-massive MBH mergers detected by \emph{LISA} will be ideal candidates for our approach. Auspiciously, these binaries and mergers will have relatively large gravitational wave error volumes, and are thus the exact scenarios in which the addition of our approach is most helpful. For low-mass and low-mass ratio MBH mergers at high redshifts (such as the majority of \emph{LISA} detections), our approach will be less useful.

We emphasize that our morphology-based approach should be used \emph{in addition} to standard redshift and stellar mass cuts on galaxies in gravitational wave localization regions. Our results suggest that morphological approaches will be useful only after those first cuts are performed and we are left with a small handful of candidate host galaxies. If the resulting number of candidate host galaxies that pass these first cuts is large (e.g., $>$10), the $\gtrsim$80\% accuracies achieved through our approach will not be able to narrow down these candidates to a single galaxy, and thus additional approaches will be needed. Nevertheless, in the absence of transient or variable electromagnetic emission stemming from the merger, our morphological approach will be one of the only methods to narrowing down candidate host galaxies.

Finally, we note that morphology-based approaches to identifying MBH merger host galaxies may only be efficacious using imaging with high resolution and signal-to-noise ratios. In this work, we assumed pixel scales and PSFs approximately matched to design specifications for future wide-field space-based telescopes such as \emph{CASTOR} and \emph{Roman}. 
For observed images with specifications similar to these assumptions, our trained LDA predictor in Equation~\ref{eq:LD1selection} can be applied directly. However, our results will likely change for ground-based imaging (e.g., with the Vera Rubin Observatory), since galaxy morphologies will be more difficult to measure with larger PSFs.

\section{Conclusions} 
\label{sec:conclusions}

We investigate the efficacy of using galaxy morphology to identify the host galaxies of MBH mergers, by studying mock galaxy images from the Romulus25 cosmological simulation of galaxy formation. We selected a sample of simulated MBH merger host galaxies with black hole masses of $M_\text{BH} \gtrsim 10^7$ and mass ratios of $\nicefrac{M_2}{M_1} > 0.1$, along with a mass- and redshift-matched control sample of galaxies not hosting MBH mergers. We generate mock broadband UV, optical, and IR images of these galaxies using the \texttt{Powderday} stellar population synthesis and dust radiative transfer simulation. We measure seven morphological statistics from the mock images: $Gini$, $M_{20}$, concentration ($C$), asymmetry ($A$), smoothness ($S$), shape asymmetry ($A_S$), and Sérsic index ($n$). We combine these measurements using linear discriminant analysis (LDA) into a linear classification function of the morphological statistics which maximally discriminates between the MBH merger hosts and control sample. Our main conclusions are:

\begin{enumerate}
    \item Galaxy morphology in UV, optical, and IR imaging can be used to identify the host galaxies of MBH mergers, using combination of multiple morphological statistics in a linear discriminant analysis. The accuracy of this approach significantly increases with chirp mass and mass ratio. At high chirp masses ($\gtrsim$10$^{8.2}$ $M_\odot$) and high mass ratios ($\gtrsim$0.5), the accuracy reaches $\gtrsim$80\%. 
    \item In contrast to previous studies, the discriminating power of our morphology-based approach does not decline for at least $\sim$1~Gyr after the numerical MBH merger, and will likely remain high when gravitational waves are detected. This is because our approach does not explicitly search for morphological disturbances from the preceding galaxy merger (which are short-lived), and is instead also sensitive to permanent morphological signatures of MBH binary and merger host galaxies.
    \item The dominant morphological signature of MBH mergers is the presence of a classical bulge, rather than morphological disturbances from galaxy mergers. Our simulated MBH merger host galaxies indeed have more prominent bulges in comparison to the mass- and redshift-matched control sample. These prominent classical bulges are formed from major mergers of massive galaxies, which then act as permanent signposts in the host galaxies of MBH binaries and mergers detected in gravitational waves. This explains why the LDA accuracy (\emph{a}) remains constant over time, (\emph{b}) increases with chirp mass and mass ratio, (\emph{c}) is higher at optical and IR wavelengths than UV, and (\emph{d}) slightly decreases with redshift.
    \item Our morphology-based approach is most useful for MBH binaries detected by PTAs, which will have high chirp masses and mass ratios, and low redshifts. If \emph{LISA} is also able to detect such systems near the low-frequency edge of its sensitivity curve, our approach can also be applied using high-resolution follow-up imaging with space-based telescopes.
\end{enumerate}

Our results suggest that morphological approaches to identifying MBH merger host galaxies may play a promising role in future telescope follow-up of MBH mergers detected in lower-frequency gravitational waves. When PTAs or \emph{LISA} detects a high mass and high mass ratio MBH merger, conventional redshift and mass selection cuts to galaxies lying in the gravitational wave localization region will first narrow down the candidate host galaxies by orders of magnitude, but identifying the exact host galaxy in the remaining small handful of candidates will be challenging. Follow-up imaging of the localization region with space-based telescopes such as \emph{CASTOR} and \emph{Roman} will then enable morphology-based identification of the exact host galaxy among the remaining candidates. Ultimately, these combinations of multi-messenger observations of MBH mergers will yield samples of host galaxies that enable studies of MBH merger environments, cosmology, and galaxy evolution.

\begin{acknowledgments}

We thank Thomas R.\ Quinn, Jessie Runnoe, Stephen Taylor, Kelly Holley-Bockelmann, and Maria Charisi for insightful discussions.

This work made extensive use of the \href{https://docs.computecanada.ca/wiki/Cedar}{Cedar} cluster of \href{https://www.computecanada.ca/home/}{Compute Canada} at Simon Fraser University, with regional partner \href{https://www.westgrid.ca/}{WestGrid}.

Romulus25 is part of the Blue Waters sustained-petascale computing project, which is supported by the National Science Foundation (awards OCI-0725070 and ACI-1238993) and the state of Illinois. Blue Waters is a joint effort of the University of Illinois at Urbana-Champaign and its National Center for Supercomputing Applications.

J.B.\ acknowledges support from the NSERC Summer Undergraduate Research Award program. J.J.R.\ and D.H.\ acknowledge support from the Canada Research Chairs program, the NSERC Discovery Grant program, the FRQNT Nouveaux Chercheurs Grant program, and the Canadian Institute for Advanced Research. J.J.R.\ acknowledges funding from the Canada Foundation for Innovation, and the Qu\'{e}bec Ministère de l’\'{E}conomie et de l’Innovation. MT was supported by an NSF Astronomy and Astrophysics Postdoctoral Fellowship under award AST-2001810. 
 
\end{acknowledgments}

\software{
\href{https://github.com/dnarayanan/powderday}{\texttt{Powderday}}: \cite{Narayanan_2021};
\href{https://github.com/pynbody/pynbody}{\texttt{Pynbody}}: \cite{Pynbody};
\href{https://github.com/pynbody/tangos}{\texttt{Tangos}}: \cite{Tangos};
\href{https://github.com/hyperion-rt/hyperion}{\texttt{Hyperion}}: \cite{Robitaille_2011};
\href{https://github.com/yt-project/yt}{\texttt{yt}}: \cite{Turk_2011};
\href{https://github.com/cconroy20/fsps}{\texttt{FSPS}}: \cite{Conroy_Gunn_2010};
\href{https://github.com/vrodgom/statmorph}{\texttt{StatMorph}}: \cite{Rodriguez-Gomez_2019};
\href{https://scikit-learn.org}{\texttt{sklearn}}: \cite{scikit-learn};
\href{https://github.com/N-BodyShop/changa}{\texttt{ChaNGa}}: \cite{Menon_2015};
\href{http://popia.ft.uam.es/AHF/}{\texttt{Amiga Halo Finder}}: \cite{Knollmann_2009};
\href{https://www.astropy.org/}{\texttt{astropy}}: \cite{astropy18}
}

\bibliography{paper}{}
\bibliographystyle{aasjournal}

\end{document}